\documentclass[journal]{IEEEtran}
\usepackage{enumitem}
\usepackage{multirow}
\usepackage{textcomp}
\usepackage{color}
\usepackage{booktabs}
\usepackage{threeparttable}

\usepackage{cite}
\ifCLASSINFOpdf
  \usepackage[pdftex]{graphicx}
\else
  \usepackage[dvips]{graphicx}
\fi

\usepackage{amsmath, amsthm, amssymb, amsfonts}
\usepackage{mathrsfs}
\usepackage[ruled, linesnumbered]{algorithm2e}  
\usepackage{algpseudocode}
\usepackage{array}

\ifCLASSOPTIONcompsoc
  \usepackage[caption=false,font=normalsize,labelfont=sf,textfont=sf]{subfig}
\else
  \usepackage[caption=false,font=footnotesize]{subfig}
\fi

\newtheorem{definition}{Definition}

\begin{document}
%
\title{NVIF: Neighboring Variational Information Flow for Large-Scale Cooperative Multi-Agent Scenarios}
%
%
%

\author{Jiajun~Chai, ~\IEEEmembership{Student Member,~IEEE,}
        Yuanheng~Zhu,~\IEEEmembership{Senior Member,~IEEE,}
        and~Dongbin~Zhao,~\IEEEmembership{Fellow,~IEEE}

\thanks{This work was supported in part by the National Key Research and Development Program of China under Grant 2018AAA0102404, in part by the Strategic Priority Research Program of Chinese Academy of Sciences under Grant No. XDA27030400, and also in part by the National Natural Science foundation of China under Grant 62136008.}
\thanks{J. Chai, Y. Zhu, and D. Zhao are with the State Key Laboratory of Management and Control for Complex Systems, Institute of Automation, Chinese Academy of Sciences, Beijing 100190, China, and are also with the School of Artificial Intelligence, University of Chinese Academy of Sciences, Beijing 100049, China.}
}

%
%

\markboth{Journal of \LaTeX\ Class Files,~Vol.~1, No.~1, January~2022}%
{Chai \MakeLowercase{\textit{et al.}}: Cooperative Large-Scale Multi-Agent Reinforcement Learning via Efficient Neighboring Communication}
%



\maketitle

\begin{abstract}
Communication-based multi-agent reinforcement learning (MARL) provides information exchange between agents, which promotes the cooperation. However, existing methods cannot perform well in the large-scale multi-agent system. In this paper, we adopt neighboring communication and propose a Neighboring Variational Information Flow (NVIF) to provide efficient communication for agents. It employs variational auto-encoder to compress the shared information into a latent state. This communication protocol does not rely dependently on a specific task, so that it can be pre-trained to stabilize the MARL training. Besides. we combine NVIF with Proximal Policy Optimization (NVIF-PPO) and Deep Q Network (NVIF-DQN), and present a theoretical analysis to illustrate NVIF-PPO can promote cooperation. We evaluate the NVIF-PPO and NVIF-DQN on MAgent, a widely used large-scale multi-agent environment, by two tasks with different map sizes. Experiments show that our method outperforms other compared methods, and can learn effective and scalable cooperation strategies in the large-scale multi-agent system. 
\end{abstract}


\IEEEpeerreviewmaketitle

\section{Introduction}
\IEEEPARstart{M}{ulti-agent} reinforcement learning (MARL) employs reinforcement learning to solve the multi-agent system problems. There are a lot of previous works address the cooperative scenarios, such as controlling robot swarms with limited sensing capabilities \cite{Cao2013a, Jiang2019}, mastering multi-agent coordination \cite{Tang2018b, zhu2019lmi, zhu2018control} and micro-management task in real-time strategy (RTS) games \cite{Shao2019, chai2021unmas, Tang2019}, and so on. However, due to the intrinsic problem of multi-agent learning like huge state and action space, credit assignment, and communication efficiency, the MARL algorithms can hardly perform cooperation in large-scale multi-agent system. 

To achieve cooperation in multi-agent system, several methods adopt the \emph{centralized training with decentralized execution} (CTDE) framework \cite{Rashid2018}, which is a comprise between independent learning and centralized learning. It provides local autonomy to agents by decentralized execution and avoids the problem of the non-stationary environment by centralized training. However, these methods perform poorly in large-scale multi-agent tasks. Another thread of work is \emph{communication-based method}, which exchanges local information between agents according to a communication protocol to help decision-making. Some works provide the agents with discrete communication channels \cite{foerster2016, kim2019learning}, while the others provide continuous channels \cite{sukhbaatar2016learning}. However, most communication-based methods need to learn the protocol with the training process of MARL. These online-learned protocols strongly depend on the training tasks and lack the multi-task scalability. 

Large-scale MARL faces several major challenges, one of which is the problem of huge amount of information exchange between agents, which leads to the \emph{information redundancy} \cite{ding2020advances}. Some works address this problem by providing a specific graph structure for communication, rather than each pair of agents communicating with each other. Some use a learned graph structure \cite{jiang2018learning, jiang2019graph, niu2021multi}, while others use a specific rule-based graph like neighboring communication \cite{luo2019multi, sheng2020learning}. Event-triggered communication is also proposed to communicate effectively with limited bandwidth \cite{dimarogonas2011distributed, zhu2017event, hu2021event}. Besides, some works do not provide direct communication between agents, but let agents communicate with virtual agents through mean-field approximation \cite{luo2020multiagent, yang2018mean}. Although these methods have made some progress, there is still room for improvement due to the training instability of their online-learned protocols. 

In cooperative multi-agent tasks, some works provide a team reward for the agents \cite{Sunehag2018, Rashid2018}, which introduce the problem of credit assignment. In credit assignment, the system should evaluate the contribution of each agent, whose difficulty increases significantly with the number of agents. Therefore, other works choose to provide agent-specific rewards for each agent to mitigate this problem directly \cite{zhang2018fully, lin2018efficient, liu2020multi}. However, such methods lack guidance to the team strategy and need to provide theoretical analysis to ensure cooperation. 

\subsection{Contribution}
In this article, we focus on the large-scale multi-agent reinforcement learning (MARL) methods, which provides communication between agents. The main contribution is twofold: 

1) we propose the \emph{neighboring variational information flow} (NVIF) to improve the communication efficiency. It compresses the information shared by agents into a \emph{latent state} as an auxiliary feature to enrich agent observation. Besides, we implement it by a novel network architecture, so that the agents can make use of the historical information shared by others. 

2) we combine NVIF with Proximal Policy Optimization (NVIF-PPO) and Deep Q Network (NVIF-DQN). Both of them provide agent-specific rewards instead of team reward to mitigate the problem of credit assignment. We also give a theoretical analysis to illustrate that NVIF-PPO can promote cooperation. 

Finally, we conduct experiments on two large-scale multi-agent tasks with different map sizes, and compare our methods with other related SOTA methods. The results show that NVIF-PPO can perform much better than the other experimental methods in all tasks, especially in tasks with a larger number of agents. In the large-scale maps, only NVIF-PPO can complete the task within a given timesteps. In addition, we conduct extra experiments to show that the learned policies can be better scaled to other unseen tasks.

\subsection{Related Work}
Communication-based MARL methods aim to provide information exchange between agents so that the agents can access more information than their local observations. Some methods directly provide communication between all pair of agents like DIAL \cite{foerster2016} and CommNet \cite{sukhbaatar2016learning}. However, these methods may face extreme \emph{information redundancy}, that is, the agents can hardly determine which information is more important. Therefore, TarMAC \cite{abhishek2019} and ATOC \cite{jiang2018learning} are proposed to determine the message sender or receiver. SchedNet \cite{kim2019learning} and I3CNet \cite{singh2018learning} reduces the communication traffic by training the agents when to communicate. Event-triggered communication is also employed to improve the communication efficiency like ETCNet \cite{hu2021event} and event-based ADP \cite{zhang2016event}. There are also a lot of works based on mean-field approximation to simplify the information exchange between agents like MFQ \cite{yang2018mean} and ACM \cite{zhou2021large}. 

Besides, a lot of works provide a specific graph structure for communication to alleviate this problem. Some works use the graph structure learned with the process of MARL. DGN \cite{jiang2019graph}, MAGIC \cite{niu2021multi} and MAGnet \cite{malysheva2018deep} generate an online-learned graph dynamically for communication. However, the online-learned graph may destabilize the training of MARL. Therefore, I2C \cite{ding2020advances} pre-trains a prior network to help agents determine whom to communicate with. Other works adopt rule-based graph for communication. LSC \cite{sheng2020learning} designs a hierarchical mechanism to provide a more effective graph. HAMMER \cite{gupta2021hammer} and CCOMA \cite{su2020counterfactual} adopt a centralized topology to allow a powerful central agent to communicate with the others. GraphComm \cite{yuan2021graphcomm} adopts neighboring communication for communication, which is a more reasonable graph. 

Recently, actor-critic methods have achieved satisfactory performance in MARL, especially those based on proximal policy optimization (PPO) \cite{schulman2017proximal}, which is widely used in single agent RL. IPPO \cite{de2020independent} uses PPO to train agents fully independently with team reward. In order to further explore the potential of PPO in multi-agent scenarios, MAPPO \cite{yu2021surprising} summarizes some technologies to improve the performance of IPPO. CoPPO \cite{wu2021coordinated} is proposed to promote cooperation of agents based on MAPPO, and provide some theoretical analysis. However, most of them require team rewards to guide agents, which brings the problem of credit assignment, especially in large-scale multi-agent system. Some methods provide agent-specific rewards for each agent rather than a team reward to reduce the difficulty of credit assignment. Zhang \emph{et al.} \cite{zhang2018fully} propose two fully decentralized actor-critic methods with agent-specific rewards, and provide convergence results under linear approximation. cA2C \cite{lin2018efficient} uses a centralized value network and decentralized policy to tackle the large-scale fleet management. G2ANet \cite{liu2020multi} employs graph neural networks to learn the adaptive and dynamic attention value without team rewards. However, there is still a lack of theoretical analysis to ensure that the algorithms can promote cooperation. 

\subsection{Organization}
This article is organized as follows. Section II introduces the problem formulation of MARL with communication. Section III proposes NVIF with its implementation and training algorithm, combines PPO and NVIF with agent-specific rewards, and gives the theoretical analysis of cooperation. Section IV shows the experiments and results, and analyzes the learned strategies. Finally, Section V gives the conclusion.

\section{Problem Formulation}
We consider a fully cooperative multi-agent task with partially observable environment, in which the agents communicate with each other to exchange information. This task can be defined as a tuple $\mathscr{U} = \{\mathbb{S}, \mathbb{A}, \mathbb{T}, \mathbb{O}, \mathbb{R}, n, \gamma\}$, where $\mathbb{S}$ is the global state space, $\mathbb{A}$ is the joint action space, $\mathbb{T}$ is the transition function, $\mathbb{O}$ is the joint action space, $\mathbb{R}$ is the reward function, $n$ is the number of agents, and $\gamma$ denotes the discount factor of \emph{discounted cumulative reward}: $G_{i, t} = \sum_{j=0}^{\infty}\gamma^{j}r_{i, t+j}$. 

In the interaction process between the multi-agent system and environment, the system takes the joint action $\textbf{a}_t = \{a_{1, t},...,a_{n, t}\} \in \mathbb{A}$ and gets the immediate reward $\boldsymbol{r}_t = \{r_{1, t},...,r_{n, t}\}$ from environment according to the reward function $R: \mathbb{S} \times \mathbb{U} \rightarrow \mathbb{R}$. If the environment provides team reward, there is $r_{1, t}=...=r_{n ,t}$, otherwise the rewards of each agent are independent. The introduction of team reward will lead to the problem of credit assignment, that is, the contribution of each agent needs to be evaluated, which will reduce the training speed of the method. Finally, by executing the joint action, the next global state $s_{t+1}$ is produced according to $\mathbb{T}$, which specifies Pr$(s_{t+1}|s_t, \textbf{a}_t)$. 

In the large-scale multi-agent system, the gap between the local observation of agents and the global state of the system is huge, which limits the cooperation among agents. Therefore, the system requires efficient communication between agents to promote cooperation. The affect of communication is to help the agent know more about the global state $s_t \in \mathbb{S}$ of the whole system at timestep $t$, so we define the \emph{latent state} obtained by each agent $i, (i=1...n)$ through communication as $\hat{s}_{i, t}$. It can be used as an auxiliary feature of agent decision-making. Therefore, each agent can maintain a policy $\pi(a_{i, t}|o_{i, t}, \hat{s}_{i, t})$ to better make decisions, where $a_{i, t}$ is the action, and $o_{i, t}$ is the local observation. 

We also provide the definition of the state value function and advantage estimator. The true state value function of each agent given by the global state can be defined as the expected value of the future accumulative rewards under joint policy $\boldsymbol{\pi}$:
\begin{equation}
V_i^{\boldsymbol{\pi}}(s_t) = \mathbb{E}_{\boldsymbol{\pi}}\big[r_{i, t+1} + \gamma r_{i, t+2} + ... | s_t \big]
\end{equation}
where $V_i^{\boldsymbol{\pi}}(s_t)$ is the state value of agent $i$ under the global state $s_t$. Besides, we define the agent-specific advantage estimator by the Generalized Advantage Estimator (GAE) as the definition in PPO \cite{schulman2017proximal}: 
\begin{equation}
\begin{aligned}
\label{eq:agent-adv}
&A_i^{\boldsymbol{\pi}} = \delta_{i, t} + (\gamma \lambda) \delta_{i, t+1} + ... + (\gamma \lambda)^{T-t-1}\delta_{i, T-1}\\
&where\ \ \delta_{i, t} = r_{i, t + 1} + \gamma V_i^{\boldsymbol{\pi}}(s_{t+1}) - V_i^{\boldsymbol{\pi}}(s_t)
\end{aligned}
\end{equation}
where $s_{t+1}$ is the next state by executing $\textbf{a}_t\sim \boldsymbol{\pi}(s_t)$, $\delta_{i, t}$ is the temporal difference, $\lambda$ is the hyper-parameter of GAE, and $T$ is the timestep at the end of an episode. They are abbreviated as $V_{i, t}$ and $A_{i, t}$. Since the ground-truth global state is hard to be obtained, the $s_t$ in the above two formulas is replaced by $[o_{i, t}, \hat{s}_{i, t}]$ in practice.

\section{Method}
In this section, we focus on the cooperation of large-scale multi-agent system, and propose a new method called Neighboring Variational Information Flow (NVIF). It adopts \emph{neighboring communication} to alleviate information redundancy and pre-trains a model to compress the information collection recurrently into a \emph{latent state}. This latent state can be provided to agents as an auxiliary feature to get more information about the whole system. This communication mechanism can stabilize the training of MARL and promote cooperation, which is illustrated by the given theoretical analysis. 

\subsection{Neighboring Variational Information Flow}
We adopt the neighboring communication, which provides communication only between the agent and its neighbors, to alleviate the problem of information redundancy. Besides, in a large-scale multi-agent system, the huge gap between local observation and global state makes it difficult to achieve cooperation. Therefore, we use the data compression ability of VAE to improve the communication efficiency. 

\begin{figure}[htbp]
\centering
\includegraphics[scale=0.35]{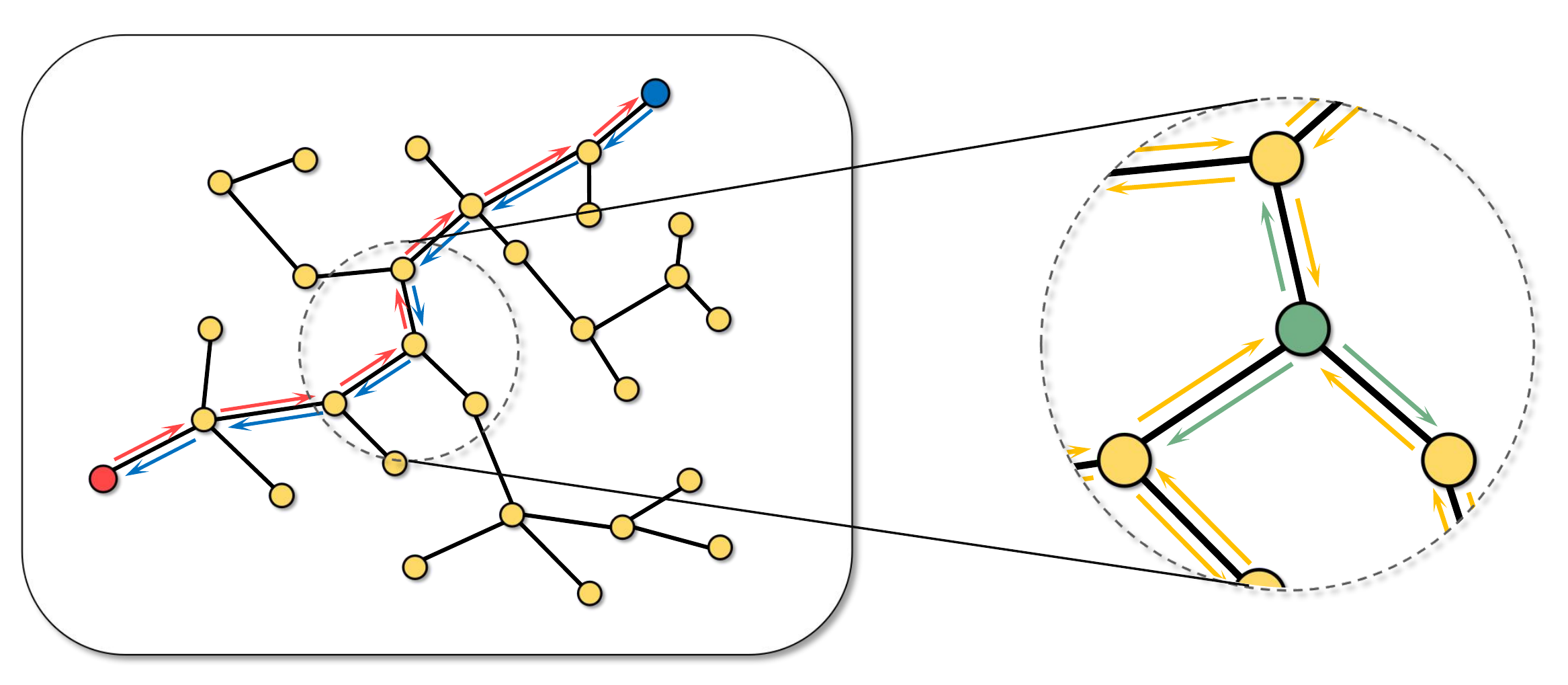}
\caption{Information flow for large-scale multi-agent system. 1) The left side shows the process of exchanging information between red agent and blue agent by neighboring communication. The shared information takes several timesteps to be received by each other. 2) The right side shows the local information flow, in which agents exchange their information to their neighbors. }
\label{fig:if}
\end{figure}

Neighboring communication is a trade-off between fully-communication and non-communication. As shown in Fig. \ref{fig:if}, the information shared by the red agent needs to take several timesteps to be accessed by the blue agent. For each agent, it can obtain the information exchanged by its neighbors, by the second-order neighbors at the last timestep, and so on. Therefore, the collection of information shared to agent $i$ at timestep $t$ can be expressed as follows:
\begin{equation}
\label{eq:info}
h_{i, t + 1} = \mathop{\bigcup} \limits_{k=t:0}\left\{\{\mathcal{I}_{j, k}\}_{j\in\beta_k}, \beta_k=\mathop{\bigcup} \limits_{j\in\beta_{k+1}}\mathcal{N}_k(j)\cup j\right\}
\end{equation}
where $h_{i, t+1}$ is the collection mentioned before, $\mathcal{I}_{i, k}$ is the information shared by agent $i$ at timestep $k$, and $\mathcal{N}_k(i)$ are its neighbors. $\beta_k$ is the collection of agents whose information shared at timestep $k$ can be accessed by agent $i$ now, where $\beta_{t+1}$ is initialized as $\{i\}$. Take $k = t$ as an example, $\beta_t = \{\mathcal{N}_t(i) \cup i\}$ is the first-order neighbors, which contain the neighbors of agent $i$ at timestep $t$ and the agent itself. Furthermore, the second-order neighbors $\beta_{t-1}$ include the neighbors of all agents in $\beta_t$ at last timestep, and so on. We name this process the \emph{neighboring information flow} to denote the flow of information in a multi-agent system. 

Furthermore, the notation of information collection $h_{i, t+1}$ can be simplified in a recurrent way as follows:
\begin{equation}
\begin{aligned}
\label{eq:info-decom}
h_{i, t + 1} 	&= \mathop{\bigcup} \limits_{k=t-1:0}\left\{\{\mathcal{I}_{j, k}\}_{j\in\beta_k}, \beta_k=\mathop{\bigcup} \limits_{j\in\beta_{k+1}}\mathcal{N}_k(j)\cup j\right\}	\\
		 	& \qquad \qquad \ \bigcup \  \{\mathcal{I}_{j, k}\}_{j\in\mathcal{N}_t(i)\cup i}\ , \ \beta_t = \mathcal{N}_t(i)\cup i 	\\
		 	&= \{h_{j, t}\}_{j\in \mathcal{N}_t(i)\cup i} \ \bigcup \ \{\mathcal{I}_{j, t}\}_{j\in\mathcal{N}_t(i)\cup i}
\end{aligned}
\end{equation}
where $h_{i, t + 1}$ is decomposed into two parts: 
\begin{enumerate}
\item Recurrent part: it contains the last information collection of the neighbors of agent $i$ and itself, and the initial value is $h_{i, 0} = \varnothing$. We define it as $\psi_{i, t} = \{h_{j, t}\}_{j\in \mathcal{N}_t(i)\cup i}$.
\item Flow part: it aggregates the information currently shared by the neighbors of agent $i$ and itself through information flow. We define it as $\varphi_{i, t} = \{\mathcal{I}_{j, t}\}_{j\in\mathcal{N}_t(i)\cup i}$.
\end{enumerate}

However, the information contained in $h_{i, t+1}$ may still not be all that the agent needs. We propose a new method called Neighboring Variational Information Flow (NVIF), which employs the VAE module to improve communication efficiency. Its encoder part compresses the information in $h_{i, t+1}$ into a latent state as the auxiliary feature for agent decision-making. In detail, the encoder part can be represented as follows: 
\begin{equation}
\hat{s}_{i, t} \sim q(s| \psi_{i, t}, \varphi_{i, t})
\end{equation}
where $q(s| \psi_{i, t}, \varphi_{i, t})$ is the inference distribution whose input is the information collection, $\hat{s}_{i, t}$ indicates the latent state output from the encoder part. 

In the decoder part, we measure the compressed latent state by reconstructing the joint observation of agents, rather than learning the communication protocols with the training of MARL. Therefore, the communication protocol trained by NVIF can be scaled to several tasks. The loss function of NVIF is modified from the loss function of VAE. It can be written as follows: 
\begin{equation}
\label{eq:loss-var}
\begin{aligned}
\mathcal{L}_{v} &= \frac{1}{n}\sum_{i=1}^n \Big[\mathbb{E}_{\hat{s}_{i, t}\sim q(s|\psi_{i, t}, \varphi_{i, t})}bce\_loss[o_{i, t}, \hat{o}_{i, t}] +  \\
&\qquad \qquad \qquad KL[q(s|\psi_{i, t}, \varphi_{i, t}) | p(s)]\Big]
\end{aligned}
\end{equation}
where $\hat{o}_{i, t}$ is the reconstructed observation output by the decoder $\mathcal{D}(\hat{s}_{i, t}, x_{i, t})$, whose inputs are the latent state $\hat{s}_{i, t}$ and the unique information $x_{i, t}$ such as position of agent $i$. It can be a binary cross entropy loss in practice. $p(s)$ is the prior distribution of latent state, which is set as the standard normal distribution $\boldsymbol{N}(\boldsymbol{0}, \boldsymbol{I})$. The first part of this loss function is the reconstruction loss of the joint observation, and the second part is the KL divergence between inference distribution and the prior distribution.

However, training directly with Eq. (\ref{eq:loss-var}) may cause the latent state to degenerate into the current local observation. Therefore, we define an additional consistency loss on the basis of Eq. (\ref{eq:loss-var}):
\begin{equation}
\label{eq:loss-con}
\mathcal{L}_{c} = \frac{1}{n}\sum_{i=1}^n \left(\hat{s}_{i, t} - \frac{1}{n}\sum_{j=1}^n\hat{s}_{j, t}\right)^2. 
\end{equation}

This loss function promotes agents to maintain the same latent state, which means that each agent can reconstruct its own local observation according to a same latent state. Therefore, it can be seen as a representation of the global state of the multi-agent system, which contains the information required by all agents. Then, the total loss can be written as:
\begin{equation}
\label{eq:loss-total}
\mathcal{L} = \mathcal{L}_{v} + \alpha \mathcal{L}_{c}
\end{equation}
where $\alpha$ is the coefficient of consistency loss. By minimizing this loss function, each agent can get the latent state with information flow. 

\subsection{Training Algorithm of NVIF}
\begin{figure}[htbp]
\centering
\includegraphics[scale=0.3]{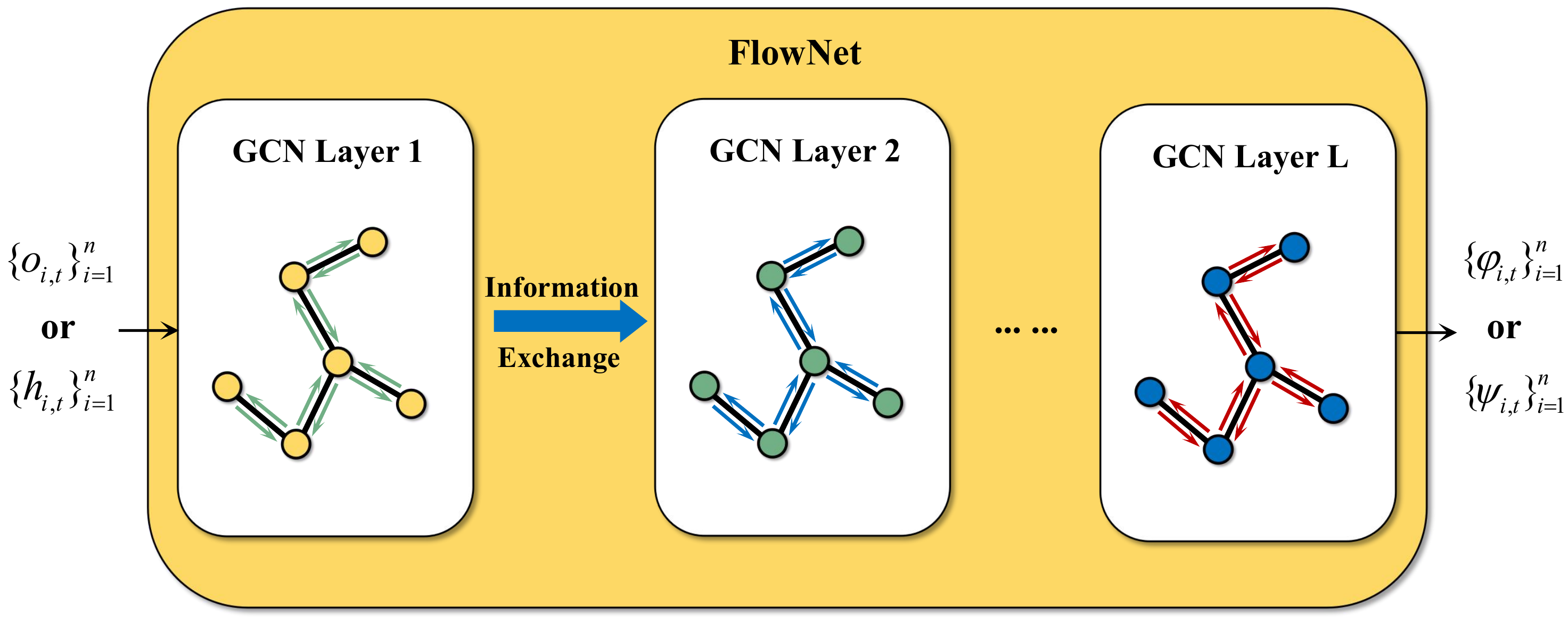}
\caption{Network architecture of FlowNet, which contains multiple GCN layers. Arrows in different colors represent the information exchanged in different layers of GCN, and circles in different colors represent the features extracted by the agent through one layer of GCN.  }
\label{fig:flownet}
\end{figure}

In this section, we propose a novel network architecture to implement NVIF and its training algorithm. As described in Eq. (\ref{eq:info-decom}), the information collection can be decomposed in a recurrent way. In the encoder part, in order to realize information exchange in the neighboring communication graph structure, we employ a Graph Convolutional Network (GCN) to simulate this process. As shown in the Fig. \ref{fig:flownet}, we refer the multi-layer GCN networks as \emph{FlowNet}. 

The GCN layer in FlowNet aims to encode both graph structure and node features of a graph through the adjacency matrix $G_t$ and original node features $H^{0}_t$. The multi-layer GCN propagation rule is:
\begin{equation}
H^{l+1}_t = ReLU\left(\tilde{B}^{-\frac{1}{2}}_t \tilde{G}_t\tilde{B}^{-\frac{1}{2}}_t H^{l}_t W^{l}\right)
\end{equation}
where $H^{l}_t$ are the features extracted by the $l$-th layer according to the local information flow of agent $i$, and $H^{0}_t = \{o_{i, t}\}_{i=1}^n$ or $\{h_{i, t}\}_{i=1}^n$. If there exists $L$ layers of GCN, then $H_t^{L+1}=\{\varphi_{i, t}\}_{i=1}^n$ or $\{\psi_{i, t}\}_{i=1}^n$. $\tilde{G}_t=G_t+I$ is the local adjacency matrix $G_t$ with self-loop $I$. $\tilde{B}_t = diag(\sum_j \tilde{G}_t[:, j])$ is a diagonal degree matrix whose elements are the sum of each row of $\tilde{A}_t$. $W^l$ is the trainable parameter. 

\begin{figure}[htbp]
\centering
\includegraphics[scale=0.35]{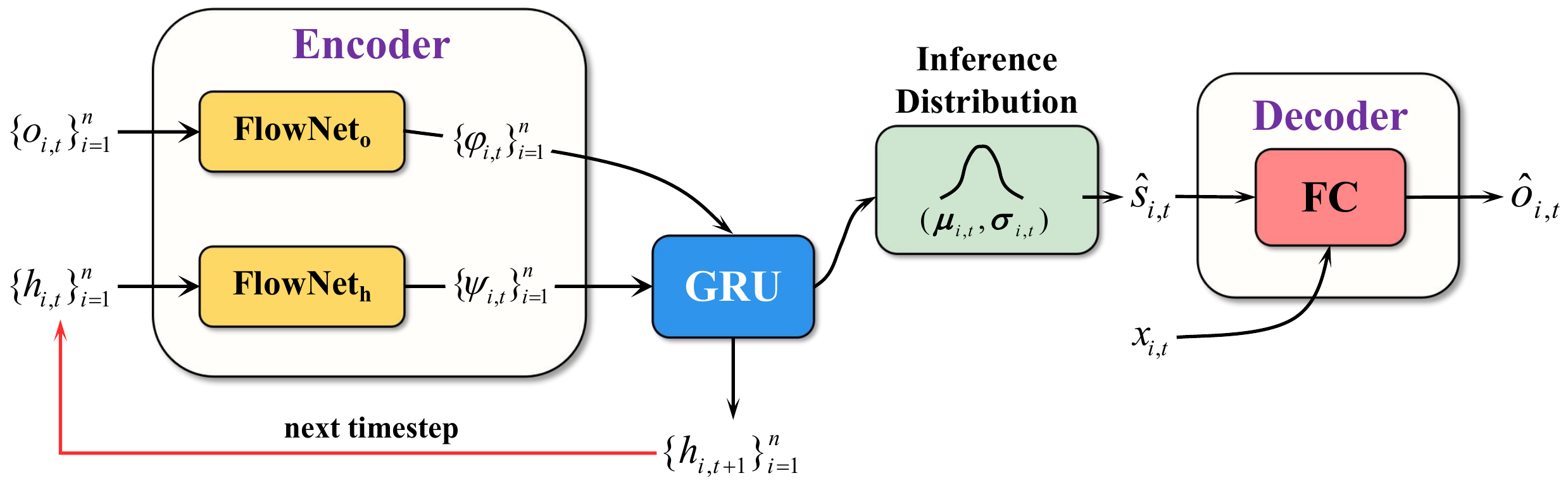}
\caption{Network architecture of NVIF. 1) The encoder uses FlowNet to simulate the information exchange process, and uses GRU to realize the recurrent property of hidden state. There are two independent FlowNets to process different inputs. 2) The decoder uses the latent state to reconstruct the observation of each agent. }
\label{fig:NVIF}
\end{figure}

The multiple layers of GCN can be seen as a case of \emph{multi-round communication}, whose rounds number is equal to the layers number. The increase of GCN layers can improve the number of information exchange in a timestep, but it will increase the training difficulty. As shown in Fig. \ref{fig:NVIF}, $FlowNet_o$ and $FlowNet_h$ take observations and hidden states as input respectively and output the corresponding extracted features:
\begin{equation}
\begin{aligned}
\{\varphi_{i, t}\}_{i=1}^n &= FlowNet_o(\{o_{i, t}\}_{i=1}^n)	\\
\{\psi_{i, t}\}_{i=1}^n &= FlowNet_h(\{h_{i, t}\}_{i=1}^n)
\end{aligned}
\end{equation}
where $FlowNet_o$ and $FlowNet_h$ process their inputs independently. 

As shown in Fig. \ref{fig:NVIF}, Besides, we introduce a Gated Recurrent Unit (GRU) layer, which is a kind of recurrent network to process sequential inputs, to realize the recurrent property of $h_{i, t+1}$: 
\begin{equation}
h_{i, t+1}, (\boldsymbol{\mu}_{i, t}, \boldsymbol{\sigma}_{i, t}) = \textbf{GRU}(\varphi_{i,t}, \psi_{i, t}). 
\end{equation}
For each agent, it takes $\varphi_{i, t}$ and $\psi_{i, t}$ extracted by FlowNets as inputs, and outputs the next hidden state. $\boldsymbol{\mu}_{i, t}$ and $\boldsymbol{\sigma}_{i, t}$ are the mean and standard deviation of the inference distribution $\boldsymbol{N}(\boldsymbol{\mu}_{i, t}, \boldsymbol{\sigma}_{i, t})$, which is a normal distribution. By sampling from the inference distribution for each agent, the information contained in $h_{i, t+1}$ are compressed into a latent state:
\begin{equation}
\hat{s}_{i, t} \sim \boldsymbol{N}(\boldsymbol{\mu}_{i, t}, \boldsymbol{\sigma}_{i, t}).
\end{equation}

\begin{algorithm}[htbp]
  \caption{Training Algorithm for Neighboring Variational Information Flow}
  \label{alg:NVIF}
  
  Collect the memory buffer $\mathcal{M}_{NVIF} = \{\boldsymbol{o}_t, \boldsymbol{x}_t, G_t \}$ using random policy\; 
  
  Initialize the parameters\; 
  \For {$episode = 1 \ to \ b$}
  {
  	Initialize the hidden state $\boldsymbol{h}_0$ as zero vectors\;
  	\While {$episode \ is \ not \ end$}
  	{
  		\For {$each \ alive \ agent \ i$}
  		{
  			Extract features $\varphi_{i, t}$ and $\psi_{i, t}$ by FlowNet\;
  			$h_{i, t+1} = \textbf{GRU}(\varphi_{i, t}; \psi_{i, t})$\;
  			$\boldsymbol{\mu}_{i, t}, \boldsymbol{\sigma}_{i, t} = f(h_{i, t+1})$\;
  			Sample latent state $\hat{s}_{i, t}$ from $\boldsymbol{N}(\boldsymbol{\mu}_{i, t}, \boldsymbol{\sigma}_{i, t})$\;
  			Reconstruct observation $\hat{o}_{i, t} = \mathcal{D}(\hat{s}_{i, t}, x_{i, t})$\;
		}
	}
  	Optimize the \emph{Encoder} and \emph{Decoder} modules with the loss function shown in Eq. (\ref{eq:loss-total})\; 
  }
  Execute the above cycle until convergence.
  
  Take the Encoder module as the final model for the information flow. 
\end{algorithm}

In the decoder part, we choose the position of agent as the unique information $x_{i, t}$, and concat it and the latent state. As shown in Fig. \ref{fig:NVIF}, we use a fully-connected (FC) network as the decoder part to get the reconstructed observation:
\begin{equation}
\hat{o}_{i, t} = \mathcal{D}(\hat{s}_{i, t}, x_{i, t})
\end{equation}
where $\hat{o}_{i, t}$ is the reconstructed observation of agent $i$, $x_{i, t}$ is its position, and $\mathcal{D}(\cdot)$ is the decoder network. 

As shown in Algorithm \ref{alg:NVIF}, we propose a training algorithm to train NVIF. The memory buffer $\mathcal{M}_{NVIF}$ contains the observations $\boldsymbol{o}$, positions $\boldsymbol{x}$ of agents, and the adjacency matrix $G$ of the system. It is collected by executing random polices and interacting with the environment. In the information exchange phase, the FlowNet of encoder part takes the observations $\{o_{i, t}\}_{i=1}^n$, hidden state $\{h_{i, t}\}_{i=1}^n$, and the adjacency matrix $G$ as inputs. After several rounds of communication, it outputs two features $\varphi_{i, t}$ and $\psi_{i, t}$. Then, the GRU layer is employed recurrently to generate the next hidden state $h_{i, t + 1}$ and the inference distribution $\boldsymbol{N}(\boldsymbol{\mu}_{i. t}, \boldsymbol{\sigma}_{i, t})$. Finally, each agent samples a latent state $\hat{s}_{i, t}$ from it. 

The decoder part reconstructs the observation of agent $i$ according to its position $x_{i, t}$ and the latent state $\hat{s}_{i, t}$. After calculating the reconstruction $\hat{o}_{i, t}$ of all agents at all times in an episode, we optimize the parameters of NVIF with the loss function shown in Eq. (\ref{eq:loss-total}). After traversing all episodes in $\mathcal{M}_{NVIF}$, the training completes an epoch. Then we repeatedly train several epochs until the loss function converges. 

After the training of NVIF, we use its encoder part to achieve our goal, so that each agent compress a large amount of information into the latent state. This model is pre-trained, which means that it can provide stable communication message for the training of MARL and does not rely strongly on a specific task. 

\subsection{Multi-Agent Reinforcement Learning with NVIF}
Since the latent state can be seen as a representation of the global state, we can use it to enrich the observation of agents and promote cooperation. Since adopting team reward for agents will face the problem of credit assignment, we combine NVIF and PPO to train agents in the large-scale multi-agent system by providing agent-specific rewards. We refer this algorithm as NVIF-PPO. 

In Algorithm \ref{alg:mappo}, We adopt the framework of PPO to train each agent. It provides two networks for each agent: an actor network presenting its policy, and a critic network presenting its value function. Both networks contain only two layers of fully-connected networks, and all agents share the same parameters. 

In our method, the communication protocol is pre-trained before the training of MARL, so that the agents can maintain a stable information exchange process. At each timestep $t$, we use the encoder part of NVIF to collect the local observations $o_{i, t}$ and hidden states $h_{i, t}$ from the neighbors of each agent $i$ and itself, and outputs the latent state $\hat{s}_{i, t}$ and next hidden state $h_{i, t+1}$. The agents choose their actions according to $o_{i, t}$ and $\hat{s}_{i, t}$ as follows: 
\begin{equation}
a_{i, t} \sim \pi_i(a|\hat{s}_{i, t}, o_{i, t}; \theta)
\end{equation} 
where $\theta$ is the parameters of the actor network. 

Besides, the agents estimate their state values by accumulating their individual rewards instead of the team reward. Since the ground-truth global state is inaccessible, the critic network is presented as $V_i(\hat{s}_{i, t}, o_{i, t}; \phi)$, where $\phi$ is the parameters of the critic network. Then, the environment executes these actions and feeds back the agent-specific reward $r_{i, t+1}$ for each agent. 

For each episode, we store $\{\boldsymbol{o}_t, \hat{\boldsymbol{s}}_t, \boldsymbol{a}_t, \boldsymbol{p}_t, \boldsymbol{r}_t, \boldsymbol{V}_t\}$ into the replay buffer $\mathcal{M}_{ppo}$ in chronological order, where these data include all agents, such as $\boldsymbol{o}_t = \{o_{i, t}\}_{i=1}^n$. $\boldsymbol{p}_t = \{p_{i, t}\}_{i=1}^n$ is the probability that action $a_{i, t}$ is selected by policy $\pi_i(a_{i, t}|\hat{s}_{i, t}, o_{i, t}; \theta)$. $\boldsymbol{V}_t$ is used for the advantage estimator $\boldsymbol{A}_t=\{A_{i, t}\}_{i=1}^n$, which is calculated and stored into the replay buffer at the end of an episode. Furthermore, $\boldsymbol{r}_t = \{r_{i, t}\}_{i=1}^n$ is used to calculate the discounted accumulate reward: 
\begin{equation}
\xi_{i, t} = r_{i, t+1} + \gamma r_{i, t+2} + ...
\end{equation}
and $\boldsymbol{\xi}_t=\{\xi_{i, t}\}_{i=1}^n$ is also stored into the replay buffer for future calculation. 

At the end of an epoch, we train the actor network by maximizing the following objective:
\begin{equation}
\begin{aligned}
\label{eq:loss-dr}
\mathcal{L}_{ar}(\theta)	&= \hat{\mathbb{E}}_t\big[\sum_{i=1}^n\min(\rho_{i, t}A_{i, t}, clip(\rho_{i, t}, 1-\epsilon, 1+\epsilon)A_{i, t})\big]	\\
				&= \hat{\mathbb{E}}_t\big[\sum_{i=1}^n\alpha_{i, t}A_{i, t}\big]
\end{aligned}
\end{equation}
where $\rho_{i, t} = \frac{\pi_i(a_{i, t}|\hat{s}_{i, t}, o_{i, t}; \theta)}{\pi_i(a_{i, t}|\hat{s}_{i, t}, o_{i, t}; \theta_{old})}$ is the ratio, $\epsilon$ is the clip coefficient, and $\alpha_{i, t}$ is the simplified clipped ratio. 

The critic network is updated by minimizing the following loss function: 
\begin{equation}
\label{eq:loss-cr}
\mathcal{L}(\phi) = \mathbb{E}_{i, t} \big[(V_i(\hat{s}_{i, t}, o_{i, t}; \phi) - \xi_{i, t})^2\big]
\end{equation}
which is calculated by the data of all agents at all timesteps in an epoch. 

\begin{algorithm}[htbp]
  \caption{Proximal Policy Optimization with NVIF in Large-Scale Multi-Agent System}
  \label{alg:mappo}
  Initialize the parameters of the actor network $\theta$ and critic network $\phi$\; 
  Initialize the replay buffer $\mathcal{M}_{ppo}$\;
  \For {$epoch = 1 \ to \ m$}
  {
  	\For {$episode = 1 \ to \ b$}
  	{
  		Initialize the hidden state $\boldsymbol{h}_0$  as zero vectors\;
  		\For {$i = 1 \ to \ T$}
  		{
  			Collect joint observation $\boldsymbol{o}_t$ and last hidden state $\boldsymbol{h}_t$\; 
  			Get latent state $\boldsymbol{\hat{s}}_t$ by NVIF\;
  			Choose actions for agents by actor network\;
  			Get state values by the critic network\;
  			Execute joint action $\boldsymbol{a}_t$ and collect  rewards $\boldsymbol{r}_t$ for all agents\; 
  			Store $\{\boldsymbol{o}_t, \hat{\boldsymbol{s}}_t, \boldsymbol{a}_t, \boldsymbol{p}_t, \boldsymbol{r}_t, \boldsymbol{V}_t\}$ into $\mathcal{M}_{ppo}$; 
		}
  		Compute $A_{i, t}$ and $\xi_{i, t}$ for all agents\;
  		Store $\{\boldsymbol{A}_t, \boldsymbol{\xi}_t\}$ into the replay buffer\;
	}
  	\For {$Update \ times \ from \ 1 \ to \ k$}
  	{
  		Update the parameters of actor network $\theta$ by maximizing Eq. (\ref{eq:loss-dr})\;
  		Update the parameters of critic network $\phi$ by minimizing Eq. (\ref{eq:loss-cr})\;
	}
  }
\end{algorithm}

However, since the environment provides agent-specific reward for each agent instead of the team reward, we have to demonstrate that the policy trained by the loss function in Eq. (\ref{eq:loss-dr}) can achieve cooperation, that is, the agents can optimize their team reward. It should be noted that although the policies are trained by agent-specific rewards, there is still a team reward used to evaluate the multi-agent system. 
\begin{definition}
Given the team reward $R_t$ of the system, the objective used to train the policy by team reward can be defined as follows:  
\begin{equation}
\begin{aligned}
\label{eq:loss-gr}
\mathcal{L}_{tr} 	&= \hat{\mathbb{E}}\big[\sum_{i=1}^n\min(\rho_{i, t}\tilde{A}_t, clip(\rho_{i, t}, 1-\epsilon, 1+\epsilon)\tilde{A}_t)\big]	\\
					&= \hat{\mathbb{E}}\big[\sum_{i=1}^n\alpha_{i, t}\tilde{A}_t\big]
\end{aligned}
\end{equation}
where $\tilde{A}_t$ is the abbreviation of $\tilde{A}_t^{\boldsymbol{\pi}}$, which is the advantage estimator calculated by the team reward. Its definition is similar with Eq. (\ref{eq:agent-adv}):
\begin{equation}
\begin{aligned}
&A_t^{\boldsymbol{\pi}} = \delta_{t} + (\gamma \lambda) \delta_{t+1} + ... + (\gamma \lambda)^{T-t-1}\delta_{T-1}\\
&where\ \ \delta_{t} = R_{t + 1} + \gamma \tilde{V}^{\boldsymbol{\pi}}(s_{t+1}) - \tilde{V}^{\boldsymbol{\pi}}(s_t)
\end{aligned}
\end{equation}
where the value function $\tilde{V}^{\boldsymbol{\pi}}(s_t)$ presents the cumulative discount expectation team reward under the joint policy $\boldsymbol{\pi}$:
\begin{equation}
\tilde{V}^{\boldsymbol{\pi}}(s_t) = \hat{\mathbb{E}}_{\boldsymbol{\pi}}\big[R_{t+1} + \gamma R_{t+2} + ... | s_t \big].
\end{equation}
It is abbreviated as $\tilde{V}_t$. 
\end{definition}

The experimental results of MAPPO \cite{yu2021surprising} show that the above objective can achieve satisfactory performance in many common multi-agent scenarios with team reward. 

\begin{definition}
In a cooperative multi-agent task, if maximizing the agent-specific rewards is equivalent to maximizing the team reward as follows:
\begin{equation}
\label{eq:gr-dr}
R_{t} = \sum_{i=1}^n \omega_i \cdot r_{i, t+1}
\end{equation}
where $\omega_i > 0, \forall i \in [1, n]$, then the task can be called an \textbf{additive task}. 
\end{definition}

The additive tasks are very common in multi-agent environments, including StarCraft II micro-management, predator-prey and so on. In an additive task, the relationship between the value function for agent-specific reward and the value function for team reward can be formulated as follows: 
\begin{equation}
\tilde{V}_t = \sum_{i=1}^n \omega_i V_{i, t}.
\end{equation}

Similarly, the relationship of the advantage functions can be formulated as:
\begin{equation}
\tilde{A}_t = \sum_{i=1}^n \omega_i A_{i, t}.
\end{equation}

Therefore, the result that the policy gradient calculated by Eq. (\ref{eq:loss-dr}) and Eq. (\ref{eq:loss-gr}) have the same direction is equivalent to the following formula:
\begin{equation}
\begin{aligned}
(\sum_{i=1}^n\alpha_{i, t}\tilde{A}_t)\cdot(\sum_{i=1}^n\alpha_{i, t}A_{i, t}) \ge 0	\\
(\sum_{i=1}^n\alpha_{i, t}\sum_{j=1}^n\omega_j A_{j, t})\cdot(\sum_{i=1}^n\alpha_{i, t}A_{i, t}) \ge 0 
\end{aligned}
\end{equation}
$\alpha_{i, t}$ can be rewritten as a proportion $\eta_{i, t} = \frac{\alpha_{i, t}}{\sum_{j=1}^n \alpha_{j, t}}$. In the large-scale multi-agent system, since the value of $\eta_{i, t}$ ranges from 0 to 1 and $\sum_{i=1}^n \eta_{i, t}=1$, we can approximately replace $\eta_{i, t}$ by $1/n$. Therefore, the equivalence condition becomes:
\begin{equation}
(\sum_{i=1}\omega_i A_{i, t})\cdot(\sum_{i=1}^n A_{i, t}) \ge 0
\end{equation}
when $\omega_1=...=\omega_n>0$, the above formula is always true. Therefore, the policy gradient calculated by Eq. (\ref{eq:loss-dr}) and Eq. (\ref{eq:loss-gr}) have the same direction. It should be noted that the $s_t$ here is the ground-truth global state. We use the combination of $\hat{s}_{i, t}$ and $o_{i, t}$ to approximate global state in practice. The above analysis indicates that in the large-scale multi-agent system, if the environment can provide agent-specific reward for each agent and the task is an additive task, NVIF-PPO with agent-specific reward can also promote cooperation. 

Besides, we also combine NVIF with DQN, which is referred as NVIF-DQN. Similar with NVIF-PPO, NVIF-DQN also uses the latent state provided by NVIF as auxiliary features, so that we omitted its pseudo code. However, it should be noted that NVIF-DQN do not have any theoretical guarantee to promote cooperation. 

\section{Experiments}
\subsection{Experimental Setup}
\begin{figure}[htbp]
\centering
\subfloat[normal task]{
\begin{minipage}{4cm}
\centering
\includegraphics[scale=0.2]{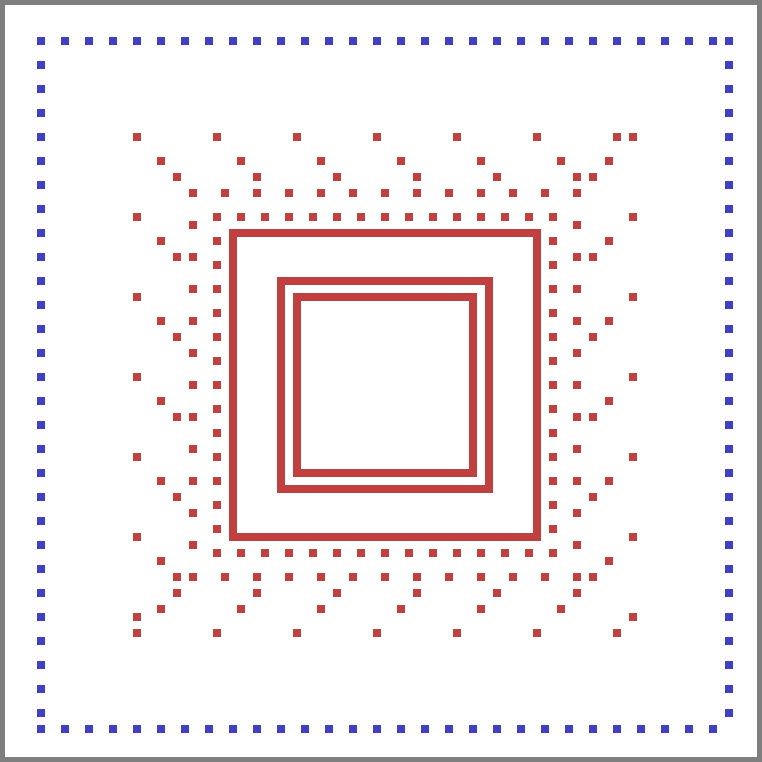}\label{fig-normal}
\end{minipage}
}
\subfloat[random task]{
\begin{minipage}{4cm}
\centering 
\includegraphics[scale=0.2]{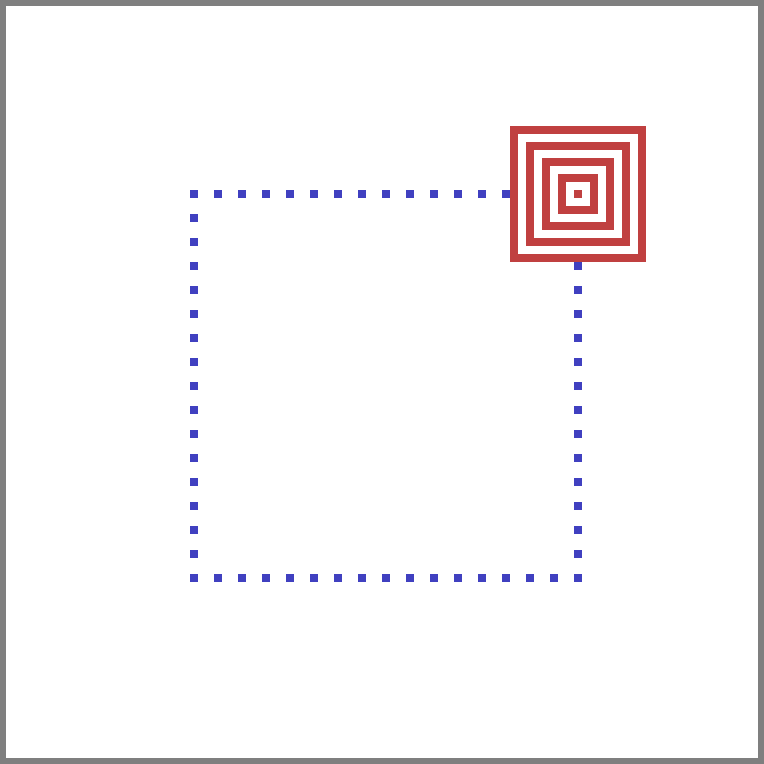}\label{fig-random}
\end{minipage}
}
\caption{\emph{Gather} game. (a) a fixed scenario, in which the initial position of agents and food is fixed. (b) the initial position of food is random, which requires efficient information exchange between agents. }
\label{fig:gather-game}
\end{figure}

MAgent\footnote{https://github.com/geek-ai/MAgent} environment \cite{zheng2018magent} is an open-source multi-agent reinforcement learning platform with large population of agents in a grid world, which is widely used by MARL methods such as MFQ \cite{yang2018mean}, DGN \cite{jiang2019graph}, and LSC \cite{sheng2020learning}. It provides appropriate interfaces to design experimental scenarios flexibly. The \emph{Gather} game shown in Fig. \ref{fig:gather-game} is a fully-cooperative multi-agent task, which contains two types of units \emph{omnivore} and \emph{food}. The blue blocks represent the omnivore units and red blocks represent the food units. Each agent controls an omnivore unit to eat food units as much as possible.

\begin{figure}[htbp]
\centering
\subfloat[observation space]{
\begin{minipage}{4cm}
\centering
\includegraphics[scale=0.3]{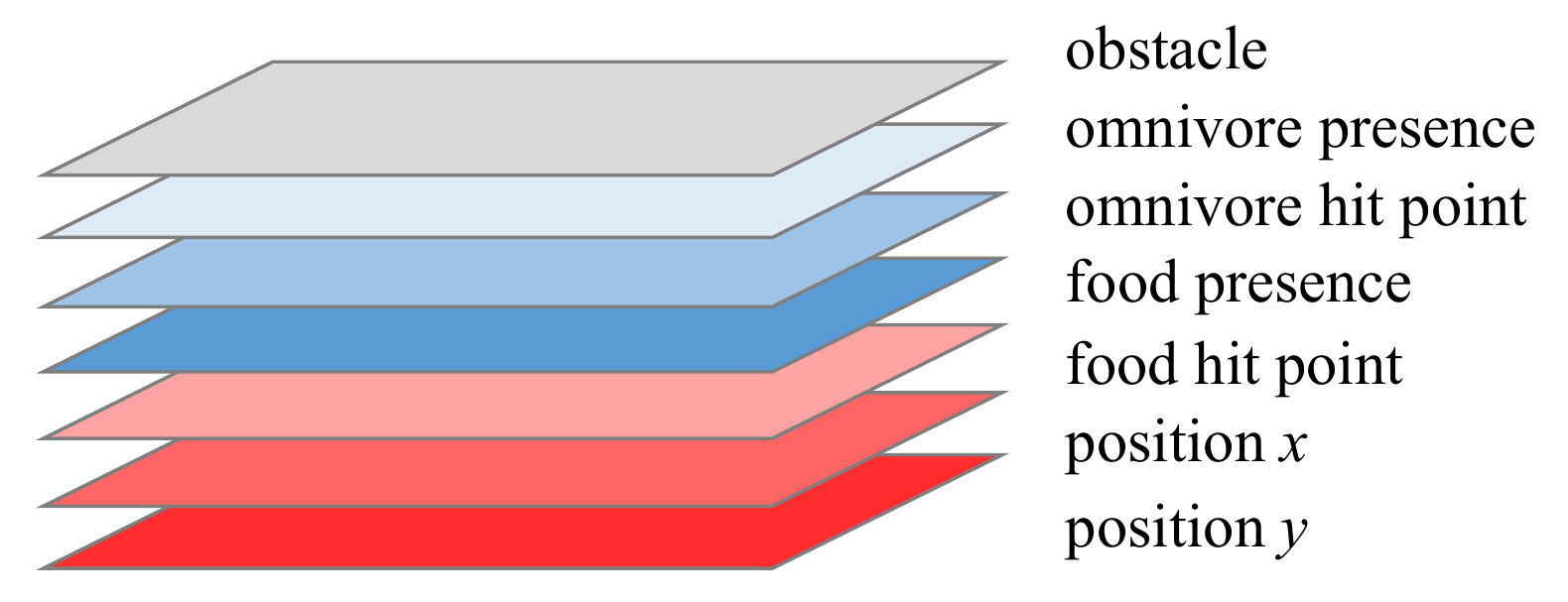}\label{fig-obs}
\end{minipage}
}
\subfloat[action space]{
\begin{minipage}{4cm}
\centering
\includegraphics[scale=0.3]{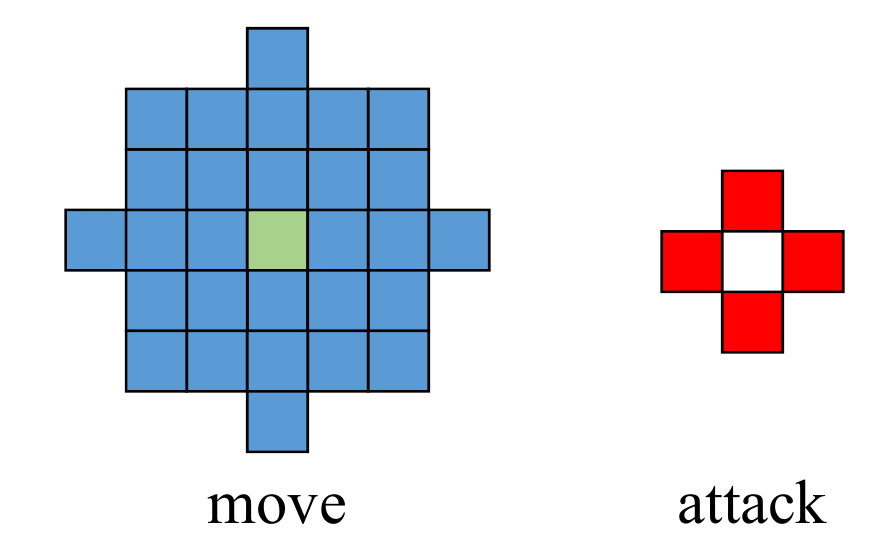}\label{fig-act}
\end{minipage}
}
\caption{The observation and action space of agent. (a) the agent observation contains all available information within the observation range. (b) the action space that contains two types of actions. }
\label{fig:obs-act}
\end{figure}

\noindent\textbf{1) Observation}: The agent observation is a local spatial view with 7 channels, which is shown in Fig. \ref{fig:obs-act} (a). The observation contains some important information of omnivores and food units within the observation range, the last two channel represent the position of the agent in the map. It should be noted that the original MAgent provides mini-map features for observations. Since it is equivalent to cheating to obtain part of the global state, we remove it in our experiments. Since the shape of the origin observation limits the information exchange process, we use a pre-trained VAE module to compress it into a one-dimensional feature. 

\noindent\textbf{2) Action}: The action space of agent contains two types of actions: \emph{move} and \emph{attack} as shown in Fig. \ref{fig:obs-act} (b). Each agent can perform 33 actions at a timestep. The blue area represents the movement range, the red area represents the attack range, and the green area means that the agent does nothing. At each timestep, the omnivore units can attack a red area within its attack range or move to any position within its movement range, while the food units keep stationary all the time. 

\noindent\textbf{3) Reward}: The reward of attack action depends on the attack target. If the target is a food unit, it will get a positive reward, and if it attacks a blank area, it will get a smaller penalty. The attack target can also be other omnivore. In this way, the attacker will get no reward, while the attacked agent will get a larger penalty and may die. Therefore, the agents should learn to eat as much food as possible without attacking and being attacked by other omnivores. Besides, there also exists a smaller penalty that encourages the multi-agent system to complete the task faster. 

\noindent\textbf{4) Tasks}: In order to evaluate the performance of proposed method, we design two types of gather tasks: \emph{normal task} and \emph{random task}. They adopt the same unit attribute settings, including hit point (HP), observation range, attack range, movement range, and so on. As shown in Fig. \ref{fig:gather-game}, in the normal task, the initial positions of all units are fixed, so that agents can make decisions only according to their positions and local observations without accessing the global state. In the random task, the initial positions of the food units are changed at the beginning of each episode. It requires efficient communication between agents, so that all agents can move forward to the correct position, that is, the position of food units. Therefore, the random task is much more difficult than the normal task, and can better evaluate the communication efficiency of the proposed method. 

\begin{figure*}[htbp]
\centering
\subfloat{
\begin{minipage}{12cm}
\centering
\includegraphics[scale=0.35]{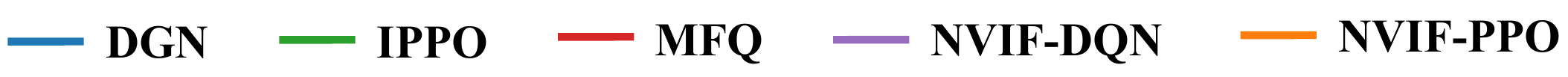}
\end{minipage}
}
\setcounter{subfigure}{0}
\subfloat[normal (small scale)]{
\begin{minipage}{5.5cm}
\centering
\includegraphics[scale=0.325]{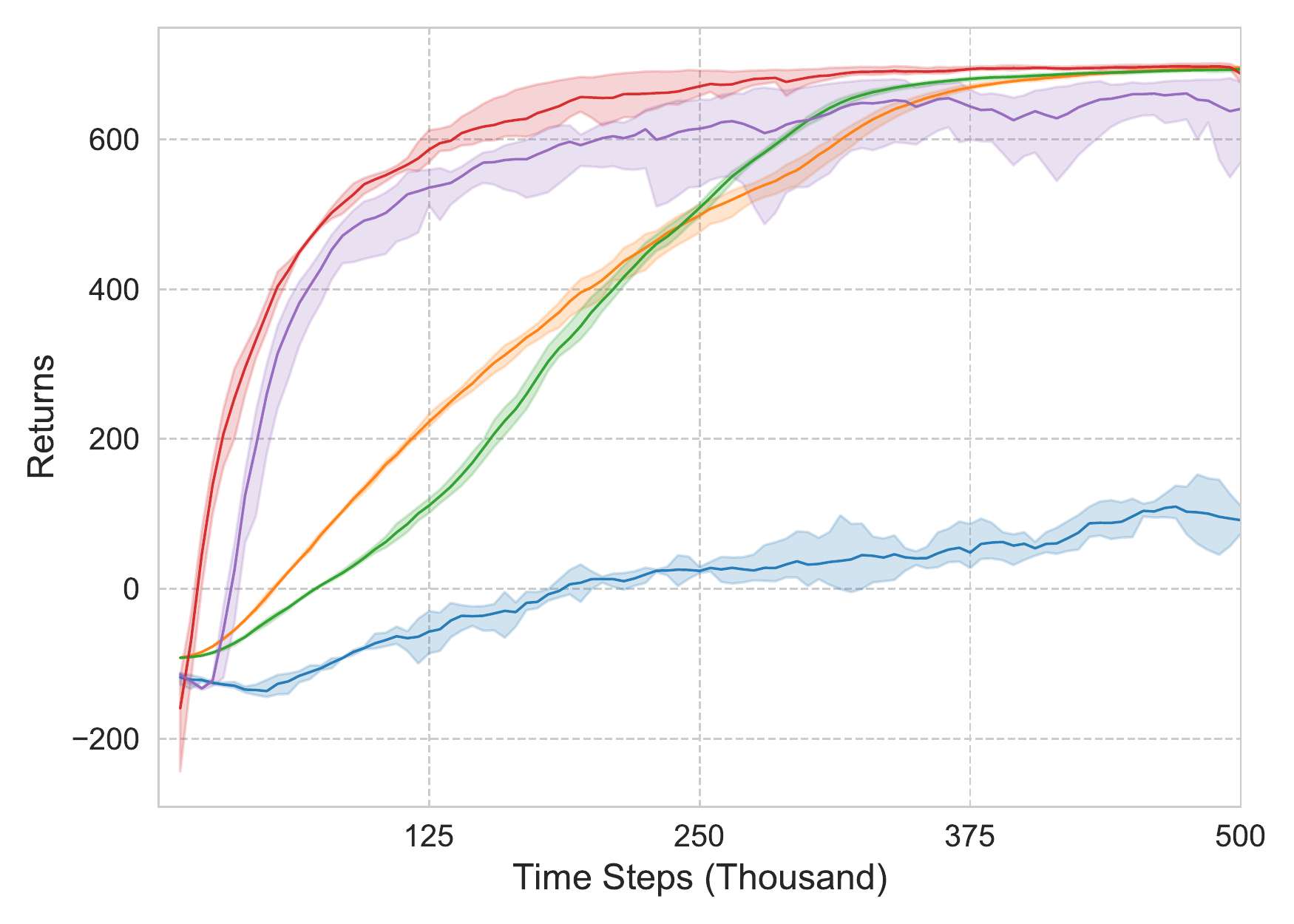}
\end{minipage}
}
\subfloat[normal (medium scale)]{
\begin{minipage}{5.5cm}
\centering
\includegraphics[scale=0.325]{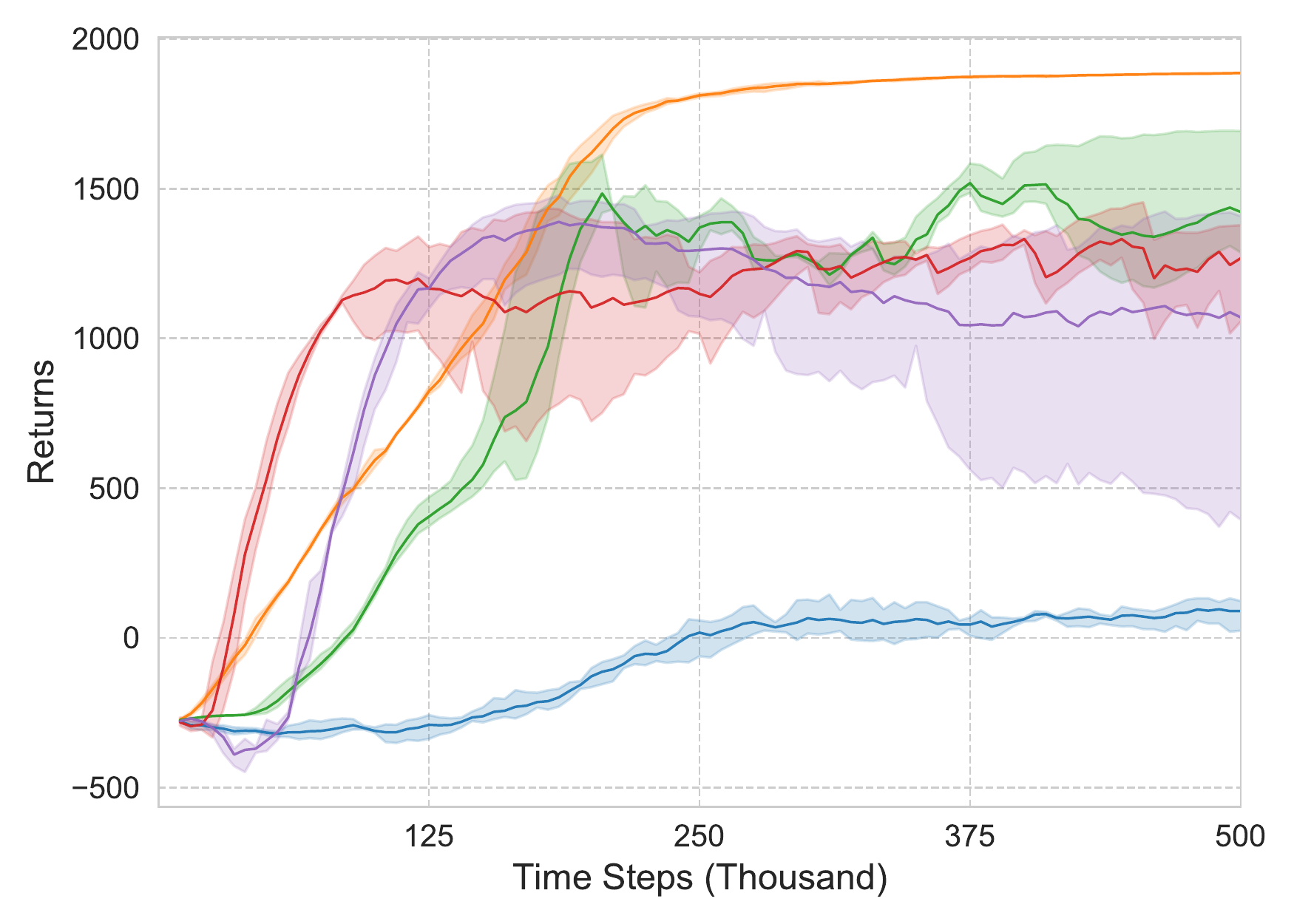}
\end{minipage}
}
\subfloat[normal (large scale)]{
\begin{minipage}{5.5cm}
\centering
\includegraphics[scale=0.325]{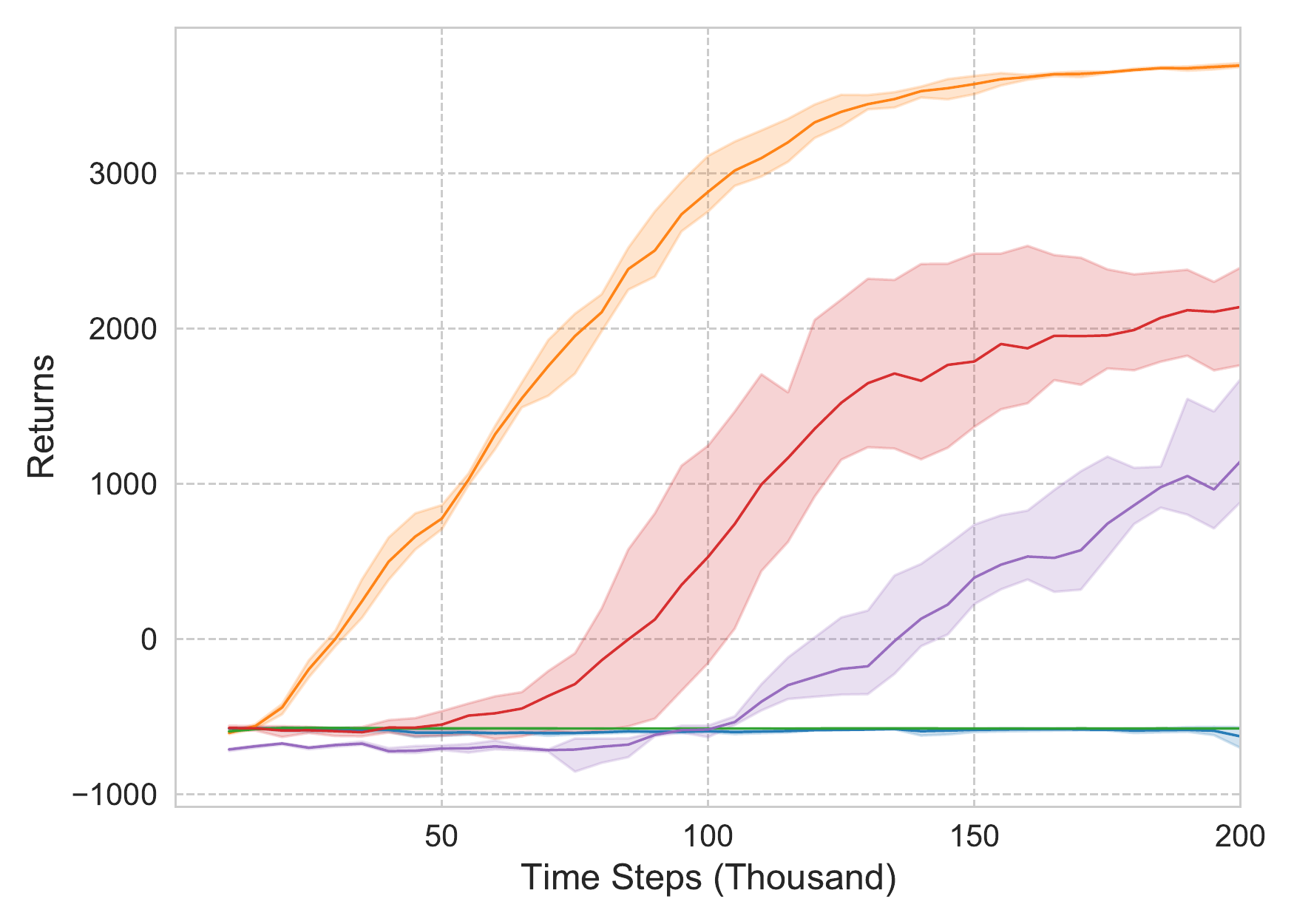}
\end{minipage}
}
\\
\subfloat[random (small scale)]{
\begin{minipage}{5.5cm}
\centering
\includegraphics[scale=0.325]{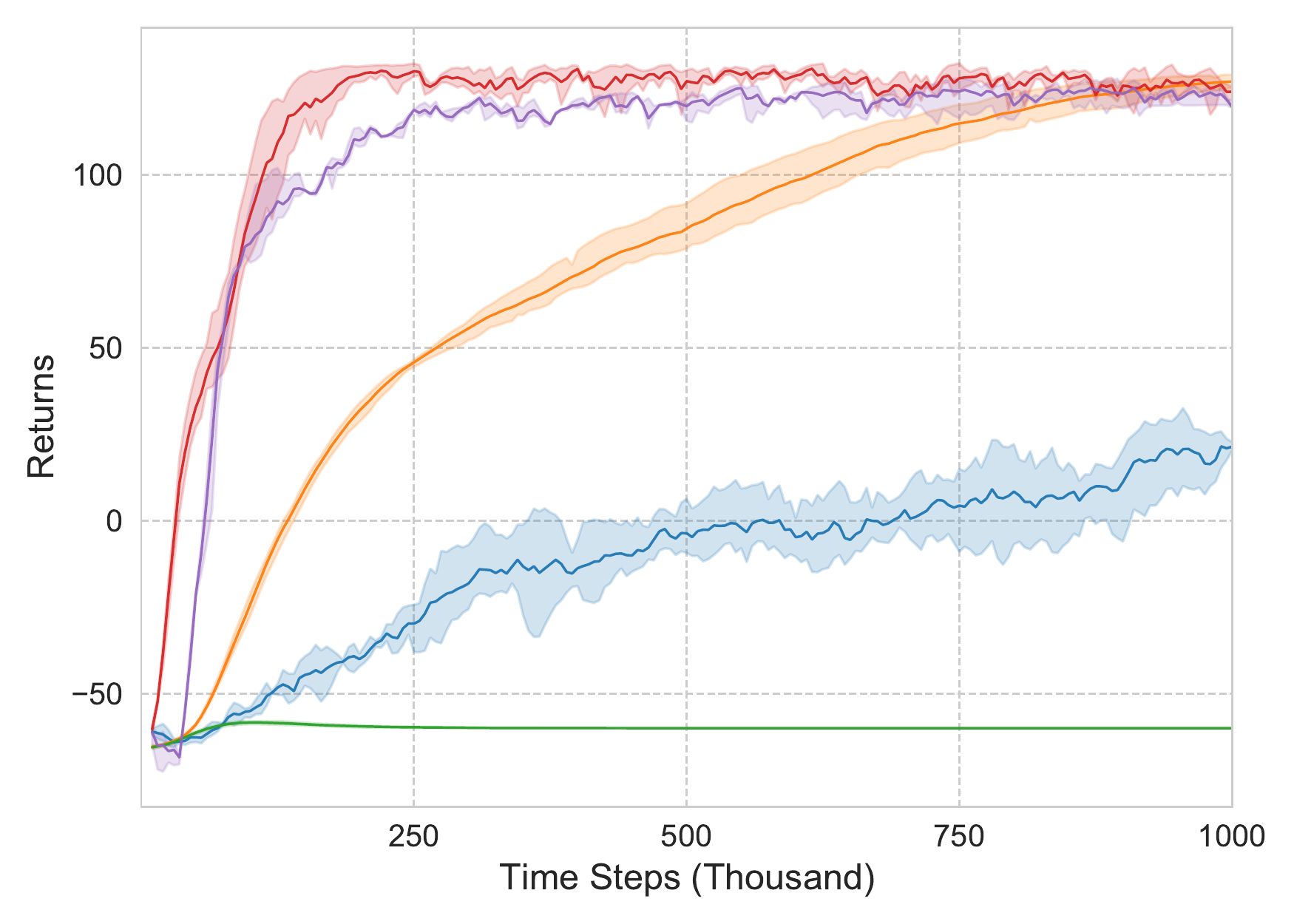}
\end{minipage}
}
\subfloat[random (medium scale)]{
\begin{minipage}{5.5cm}
\centering
\includegraphics[scale=0.325]{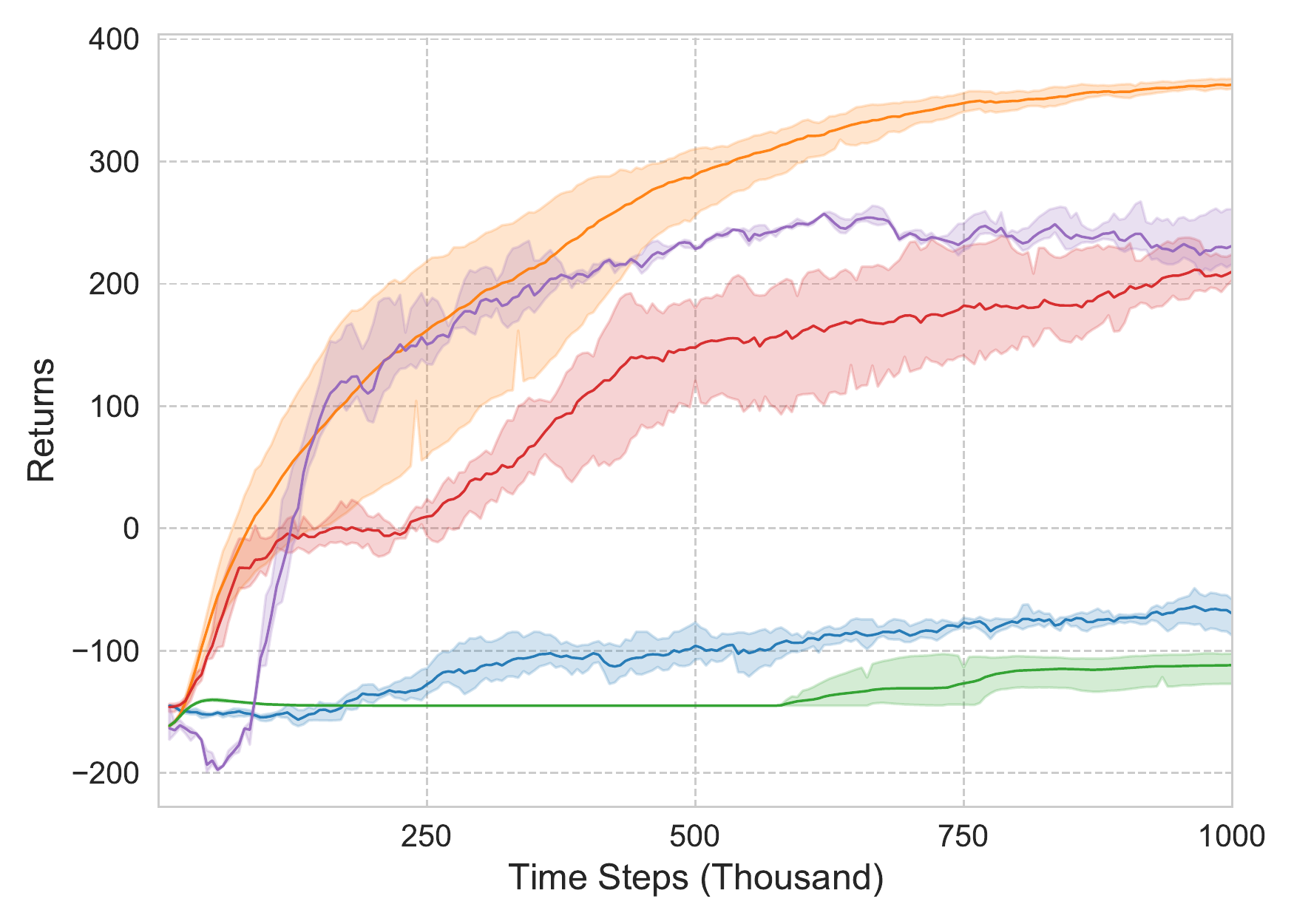}
\end{minipage}
}
\subfloat[random (large scale)]{
\begin{minipage}{5.5cm}
\centering
\includegraphics[scale=0.325]{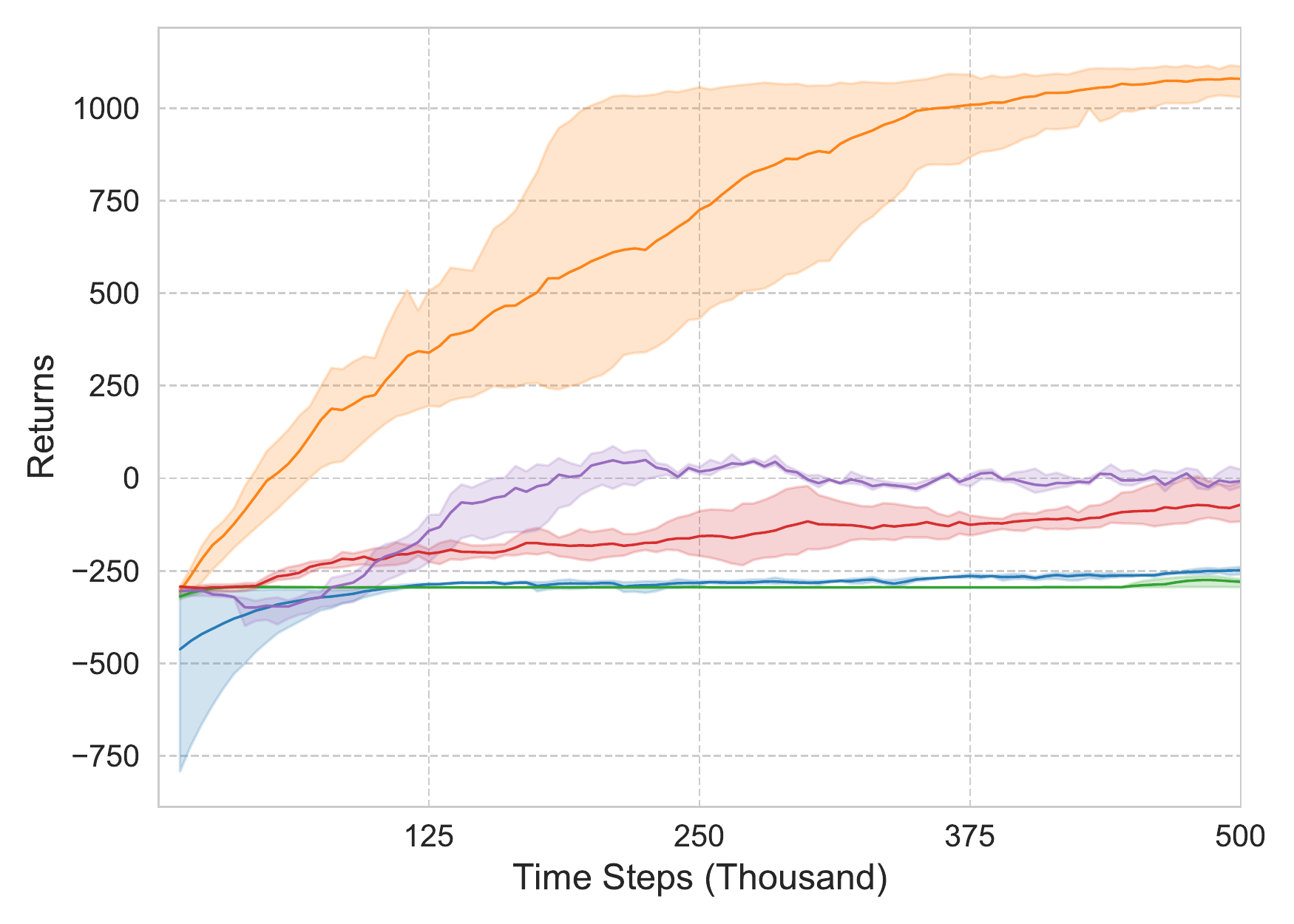}
\end{minipage}
}
\caption{Average return for DGN, IPPO, MFQ, NVIF-DQN, and NVIF-PPO. (a)(b)(c) the results in small, medium, and large-scale normal tasks. (d)(e)(f) the results in small, medium, and large-scale random tasks. The evaluation metric is the return, which is the sum of the rewards of all agents in an episode. NVIF completes all tasks and gets the best performance. }
\label{fig:result}
\end{figure*}

\begin{figure}[htbp]
\centering
\includegraphics[scale=0.3]{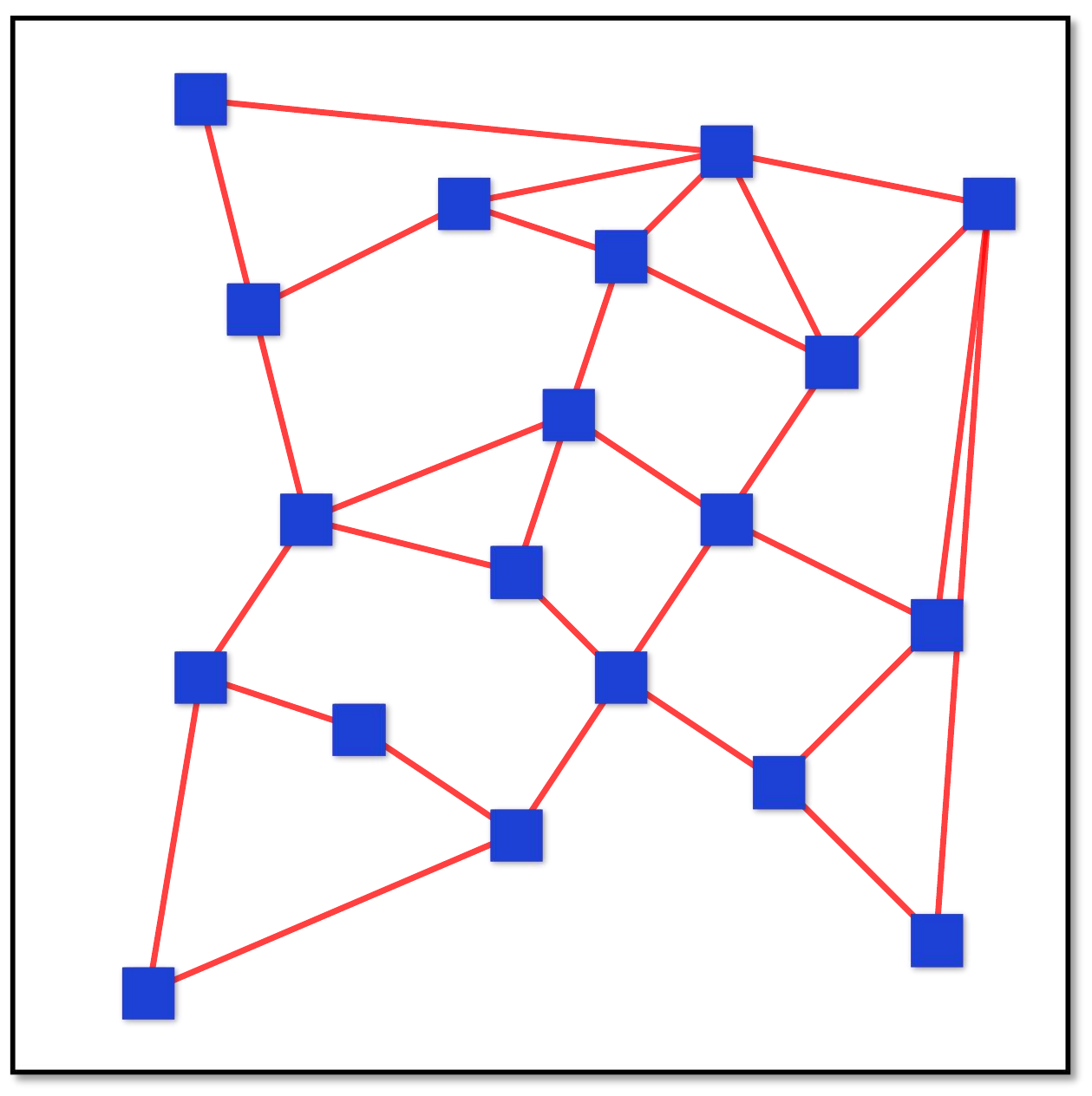}
\caption{The neighbor relationship of a multi-agent system. Each agent chooses the units with smallest distance in the up, down, right and left directions to become its neighbors. }
\label{fig:adj}
\end{figure}

Although we do not provide direct communication for each pair of agents, we should still provide an indirect communication channel between them. Therefore, the information shared by any agent can be received by all other agents. We establish a rule-based neighboring graph structure to achieve this. We define the neighbors as the units with the smallest distance from the agent in the up, down, left, and right directions as shown in Fig. \ref{fig:adj}. It should be noted that the neighbor relationship is bi-directional, and the nearest agents in two directions may be the same one. Therefore, each agent has at least one neighbor and the maximum number is unlimited. 

We compare NVIF-DQN and NVIF-PPO with MFQ \cite{yang2018mean}, DGN \cite{jiang2019graph} and IPPO \cite{de2020independent} with agent-specific rewards. MFQ \cite{yang2018mean} aims to figure out the problem of enormous interactions in the large-scale multi-agent system. It uses the mean-field approximation to learn the best response of each agent to the mean effect of its neighbors. The agents use the following Boltzmann policy to make decisions: 
\begin{equation}
\pi_i(a_{i, t}|o_{i, t}, \bar{a}_{i, t}) = \frac{\exp(-\kappa Q_i(o_{i, t}, a_{i, t}, \bar{a}_{i, t}))}{\sum_{a} \exp(-\kappa Q_i(o_{i, t}, a, \bar{a}_{i, t}))}
\end{equation}
where $\kappa$ is the temperature coefficient, $\bar{a}_{i, t}$ is the mean action of the neighbors of agent $i$, which is determined by the last mean actions. 

DGN \cite{jiang2019graph} uses the graph convolution network to provide information exchange between the multi-agent system. The agents collect the hidden states of all agents with relational kernel to alleviate information redundancy. MAPPO \cite{yu2021surprising} is an efficient MARL method, which achieves excellent performance in the StarCraft II micro-management task. However, its critic module takes the global state of system as input, which is difficult to achieve in the large scale multi-agent system. Therefore, we compare with IPPO \cite{de2020independent}, which only uses the local observation to make decisions and calculate state values for each agent, and employs PPO to update its policy. But we train it by agent-specific rewards instead of team reward. The codes of these algorithms are all open-source, and they have been conducted experiments on some tasks of MAgent. 

\begin{table}[ht]
\caption{Detailed Descriptions of Experimental Maps.}
\label{table:maps}
\renewcommand{\arraystretch}{1.1}
\setlength{\tabcolsep}{4mm}
\begin{center}
\begin{tabular}{c|cccc}
\toprule
Type	 						& Scale 			& Map Size		& Omnivores		& Food			\\ \midrule
\multirow{3}{*}{Normal}		& Small 			& 24 			& 27				& 87				\\
							& Medium 			& 48 			& 56  			& 237			\\
				 			& Large 			& 96 			& 115  			& 521			\\ \midrule
\multirow{3}{*}{Random}		& Small 			& 24 			& 15 			& 17				\\
							& Medium 			& 48 			& 29  			& 49				\\
				 			& Large 			& 96 			& 49  			& 161			\\
\bottomrule
\end{tabular}
\end{center}
\end{table}

We conduct experiments on the normal and random tasks, and provide maps of size 24, 48 and 96 to evaluate the performance of the methods under different population of multi-agent system. The detailed descriptions of maps are shown in Table \ref{table:maps}. The number of omnivore and food units increases with the size of map. In a small-scale map, the observation range is close to the size of map, so that the information exchange is inessential in the decision-making process. However, in a large-scale map, local observation loses a lot of information compared with the global state. Therefore, the agents need efficient information exchange to achieve better performance. 

\subsection{Main Results}
Since we train IPPO and our algorithm in a parallel way, while DGN and MFQ are not, we choose the average return under the same number of training timesteps as the evaluation metric to ensure the fairness of comparison. Furthermore, all algorithms use the same hyper-parameters, which have been fine-tuned to improve performance. 

\begin{enumerate}[labelsep = .5em, leftmargin = 0em, itemindent = 2em]
\item[1)] \emph{Normal Tasks}: In the normal tasks, the agents do not rely strongly on communication to obtain the position of food units, but can be obtained by training. Therefore, they should pay more attention to learn how to avoid attacking or being attacked by other agents. 

\quad As shown in Fig. \ref{fig:result} (a)(b)(c), MFQ, NVIF-DQN, and IPPO perform better on small-scale map, and achieve better convergence speed. However, with the increase of map size and agent number, their performance decreases significantly. DGN can not converge under a given number of training timesteps due to the low training speed of its attention model. NVIF-PPO can achieve the best performance on all sizes of maps. It comes from the auxiliary features provided by the pre-trained communication protocol, which makes MARL training more stable and efficient.

\item[2)] \emph{Random Tasks}: The random tasks are more difficult than the normal tasks. The initial positions of food units are random, which means that only part of agents can observe the food units at the beginning of each episode. Therefore, they need to exchange information efficiently with each other. 

\quad As shown in Fig. \ref{fig:result} (d)(e)(f), MFQ and NVIF-DQN can still perform well in small-scale map because there is less gap between the observation range and map size. However, with this gap increasing, MFQ can hardly learn a satisfactory agent policy. DGN and IPPO achieve poor performance and converge slowly. Due to the efficient communication, agents trained by NVIF-PPO can achieve the best performance in larger maps. 

\end{enumerate}

Besides, Table. \ref{table:results} shows the average returns and end timesteps of each episode when the algorithms converge. Since an episode terminates when the maximum timestep is reached or all food units are killed, less end timesteps indicates the multi-agent system has stronger cooperation ability to complete the task quickly. In general, the results of NVIF-PPO show that NVIF can provide efficient communication to help agents kill all food units in the shortest timesteps. However, due to the lack of theoretical guarantee of cooperation, the performance of NVIF-DQN is worse than that of NVIF-PPO. With the help of the latent states provided by NVIF, NVIF-DQN can achieve better performance than MFQ in random tasks, which rely more on communication, but performs worse than MFQ in normal tasks. 

\begin{table*}[ht]
\caption{Detailed Results of Experiments.}
\label{table:results}
\renewcommand{\arraystretch}{1.25}
\setlength{\tabcolsep}{2.75mm}
\begin{center}
\begin{tabular}{c|cccccc|cccccc}
\hline
\multirow{3}{*}{Method} &
  \multicolumn{6}{c|}{Normal Task} &
  \multicolumn{6}{c}{Random Task} \\ \cline{2-13} 
  &
  \multicolumn{2}{c|}{24} & \multicolumn{2}{c|}{48} & \multicolumn{2}{c|}{96} & \multicolumn{2}{c|}{24} & \multicolumn{2}{c|}{48} & \multicolumn{2}{c}{96} \\ \cline{2-13} 
  & 
  \multicolumn{1}{c|}{Return} & \multicolumn{1}{l|}{Steps} & \multicolumn{1}{c|}{Return} & \multicolumn{1}{l|}{Steps} & \multicolumn{1}{c|}{Return} & \multicolumn{1}{l|}{Steps} &
  \multicolumn{1}{c|}{Return} & \multicolumn{1}{l|}{Steps} & \multicolumn{1}{c|}{Return} & \multicolumn{1}{l|}{Steps} & \multicolumn{1}{c|}{Return} & \multicolumn{1}{l}{Steps} \\ \hline

IPPO	&	\multicolumn{1}{c|}{704.96} & \multicolumn{1}{c|}{23.68} & \multicolumn{1}{c|}{1889.83} & \multicolumn{1}{c|}{42.37} & \multicolumn{1}{c|}{401.86} & 100.00 &
  		\multicolumn{1}{c|}{132.08} & \multicolumn{1}{c|}{15.19} & \multicolumn{1}{c|}{-102.00} & \multicolumn{1}{c|}{100.00} & \multicolumn{1}{c|}{-288.69} & 100.00 \\ \hline

DGN	&	\multicolumn{1}{c|}{253.90} & \multicolumn{1}{c|}{80.00} & \multicolumn{1}{c|}{83.96} & \multicolumn{1}{c|}{100.00} & \multicolumn{1}{c|}{-192.01} & 100.00 &
  		\multicolumn{1}{c|}{42.04} & \multicolumn{1}{c|}{76.29} & \multicolumn{1}{c|}{13.83} & \multicolumn{1}{c|}{100.00} & \multicolumn{1}{c|}{-219.48} & 100.00 \\ \hline

MFQ	&	\multicolumn{1}{c|}{690.56} & \multicolumn{1}{c|}{47.79} & \multicolumn{1}{c|}{1474.53} & \multicolumn{1}{c|}{100.00} & \multicolumn{1}{c|}{2193.70} & 100.00 &
  		\multicolumn{1}{c|}{120.97} & \multicolumn{1}{c|}{28.35} & \multicolumn{1}{c|}{233.35} & \multicolumn{1}{c|}{97.99} & \multicolumn{1}{c|}{-4.42} & 100.00 \\ \hline

NVIF-DQN	&	\multicolumn{1}{c|}{639.50} & \multicolumn{1}{c|}{62.91} & \multicolumn{1}{c|}{1040.68} & \multicolumn{1}{c|}{100.00} & \multicolumn{1}{c|}{1413.48} & 100.00 &
  		\multicolumn{1}{c|}{125.24} & \multicolumn{1}{c|}{33.01} & \multicolumn{1}{c|}{234.81} & \multicolumn{1}{c|}{90.87} & \multicolumn{1}{c|}{-22.25} & 100.00 \\ \hline

NVIF-PPO	&	\multicolumn{1}{c|}{\textbf{705.27}} & \multicolumn{1}{c|}{\textbf{23.46}} & \multicolumn{1}{c|}{\textbf{1908.47}} & \multicolumn{1}{c|}{\textbf{35.36}} & \multicolumn{1}{c|}{\textbf{4108.84}} & \textbf{53.04} &
  		\multicolumn{1}{c|}{\textbf{133.51}} & \multicolumn{1}{c|}{\textbf{13.25}} & \multicolumn{1}{c|}{\textbf{379.30}} & \multicolumn{1}{c|}{\textbf{23.67}} & \multicolumn{1}{c|}{\textbf{1181.66}} & \textbf{59.57} \\ \hline

\end{tabular}
\end{center}
\end{table*}

\begin{figure}[htbp]
\centering
\subfloat{
\begin{minipage}{7cm}
\centering
\includegraphics[scale=0.35]{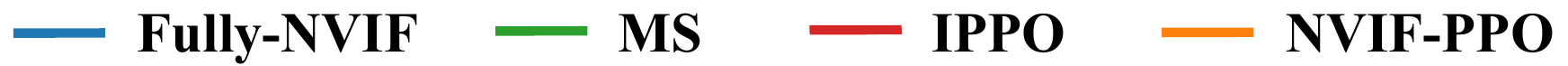} 
\end{minipage}
}
\\
\setcounter{subfigure}{0}
\subfloat[normal task]{
\begin{minipage}{7cm}
\centering
\includegraphics[scale=0.35]{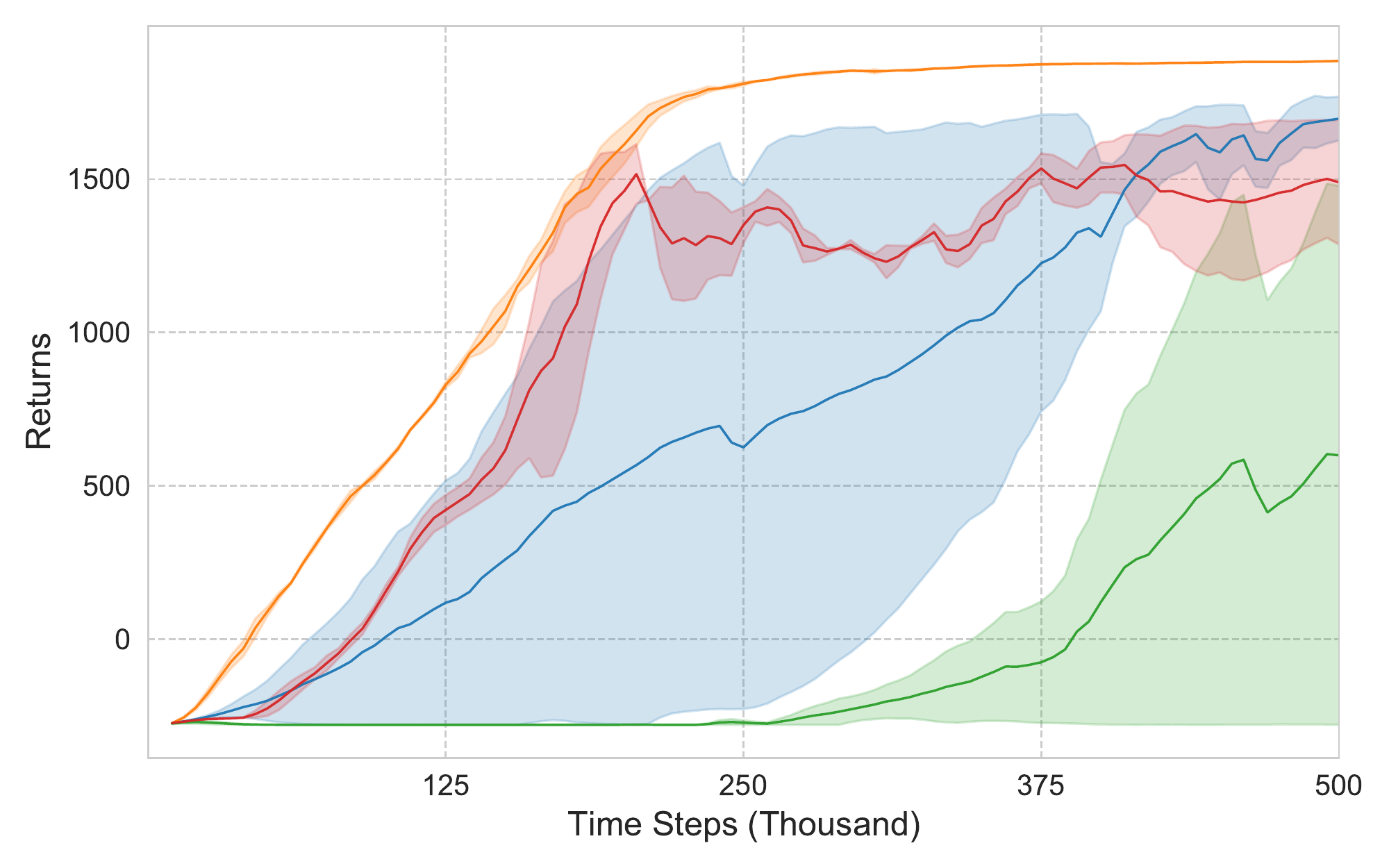} 
\end{minipage}
}
\\
\subfloat[random task]{
\begin{minipage}{7cm}
\centering
\includegraphics[scale=0.35]{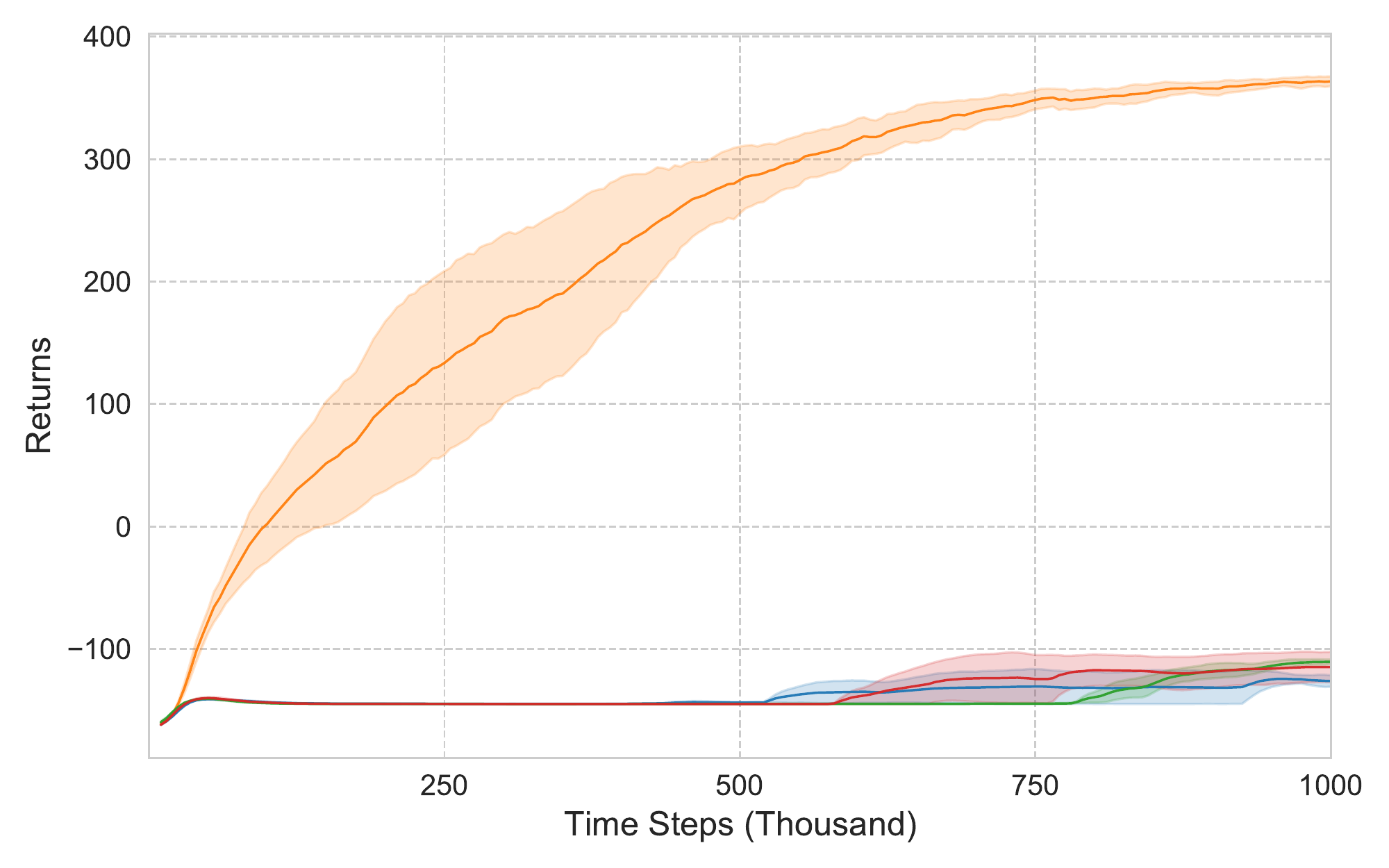}
\end{minipage}
}
\caption{Results of ablation experiments in the medium scale map of (a) normal task and (b) random task. MS takes the mean observation as the auxiliary features for agent decision making to illustrate the efficiency of NVIF. Fully-VIF uses the fully communication instead of neighboring communication to demonstrate the impact of information redundancy. }
\label{fig:ablation}
\end{figure}

\subsection{Ablation Results}
In order to investigate the effect of: 1) the neighboring communication mechanism and 2) the latent state provided by NVIF, we conduct ablation experiments on the medium scale map of the normal and random task.

\begin{enumerate}[leftmargin = 2em]
\item[1)] \emph{Effect of neighboring communication}: We propose an ablation method called Fully-VIF, which provides information exchange between each pair of agents. The neighboring communication mechanism can alleviate information redundancy, whose impact can be proved by this ablation experiment. As shown in Fig. \ref{fig:ablation}, the blue curves indicate the results of Fully-VIF. Its performance is between NVIF-PPO and IPPO, which demonstrates the necessity of neighboring communication. 

\item[2)] \emph{Effect of latent state}: We propose an ablation method called MS, which uses the mean observation of all alive agents as the auxiliary observation. It is the simplest way to obtain the information of the whole system. However, it is not as efficient and effective as the latent state provided by NVIF. As shown in Fig. \ref{fig:ablation}, the green curves indicate the results of MS. Since the average observation can not provide efficient information, it affects the training of the algorithm and leads to the decline of convergence speed. 

\end{enumerate}

According to the ablation experiments, we demonstrate the necessity of the main contribution of NVIF. The neighboring communication mechanism provides efficient information exchange by avoiding information redundancy. The VAE module of NVIF compresses the information collection into a latent state and retains the important parts. 

\subsection{Strategy Analysis}
In order to better understand the difficulties of large-scale multi-agent reinforcement learning, we analyze the replays to find out what strategies help multi-agent system achieve better performance. 

\begin{enumerate}[labelsep = .5em, leftmargin = 0em, itemindent = 2em]
\item[1)] \emph{Concentrating Attack}: It is a common strategy that can be learned by all methods. As shown in the bottom right of Fig. \ref{fig:strategy} (a), several agents move around a food unit and attack it together. Since the food units require multiple attacks to be killed, the concentrating attack can help the agents complete the task as quickly as possible.  

\begin{figure}[htbp]
\centering
\subfloat[concentrating attack \& crossing obstacles]{
\begin{minipage}{4cm}
\centering
\includegraphics[scale=0.225]{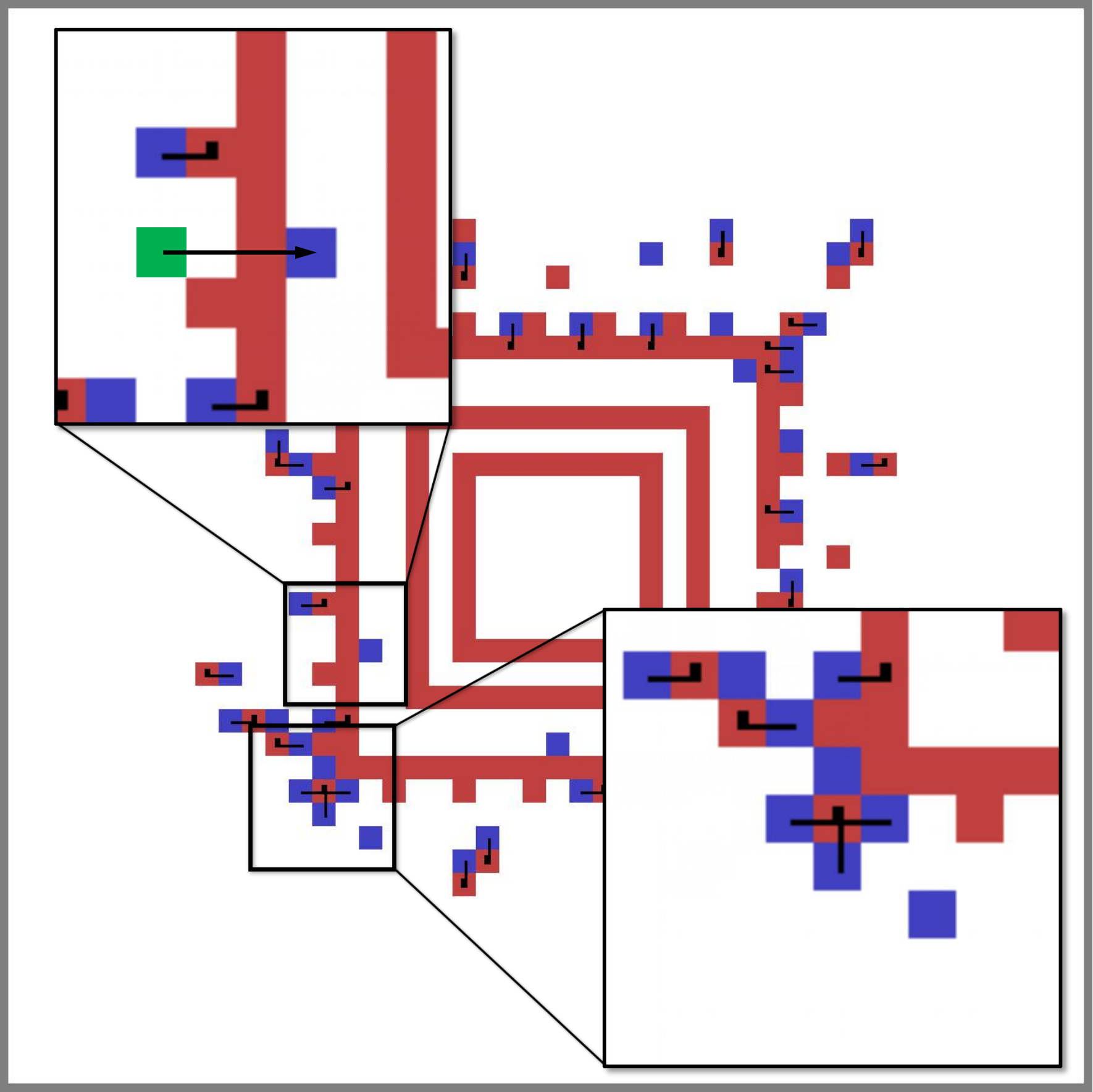} 
\end{minipage}
}
\subfloat[gathering]{
\begin{minipage}{4cm}
\centering
\includegraphics[scale=0.225]{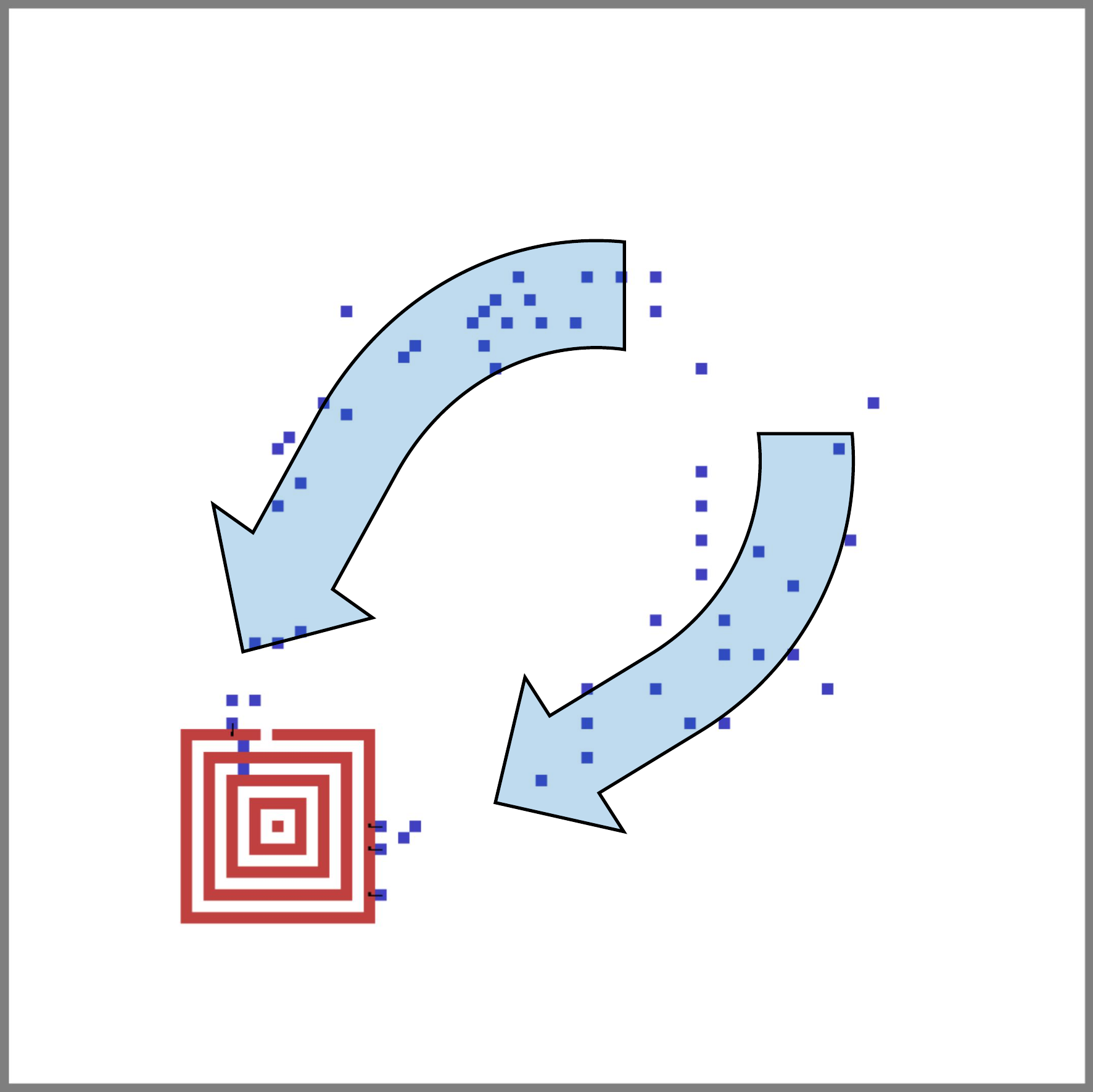}
\end{minipage}
}
\caption{Strategies learned by the agents using NVIF-PPO. There are two basic strategies \emph{concentrating attack} and \emph{crossing obstacles}, and one special strategy \emph{gathering}. (a) Top left shows the concentrating attack and bottom right shows the crossing obstacles. (b) The agents gathering directly to the correct position of food units. }
\label{fig:strategy}
\end{figure}

\item[2)] \emph{Crossing Obstacles}: It is an advanced strategy, which can further shorten the timesteps required to complete the task. As described in section IV. A, an agent can move up to 3 grids at a timestep. Therefore, the agents can move through the line made up of food units, which can be seen as an obstacle. Meanwhile, the number of food units will decrease with the attack of agents, which will lead to the aggregation of a large number of agents and cause congestion. As shown in the top left of Fig. \ref{fig:strategy} (a), the omnivore unit crosses the obstacle from the position of the green blocks, which indicates the position of the agent at the last timestep, to the end-point indicated by the black arrow. Therefore, the agent can attack the internal food units in advance to avoid the congestion and improve the attacking efficiency.

\item[3)] \emph{Gathering}: This is a special strategy learned by the agents using NVIF-PPO, which is essential in random tasks. Since the initial positions of food units are not fixed, the agents have to decide where to move according to the information exchanged by other agents. As denoted in Fig. \ref{fig:strategy} (b), all agents gather to the location of the food units. Once an agent receives the information shared from the agents who can observe the food units, it will move directly to true location. Therefore, the agents can move to the correct positions as soon as possible to complete the task. 

\end{enumerate}

\subsection{Scalability Experiments}
\begin{figure}[htbp]
\centering
\subfloat{
\begin{minipage}{0.5cm}
\centering
\includegraphics[scale=0.5]{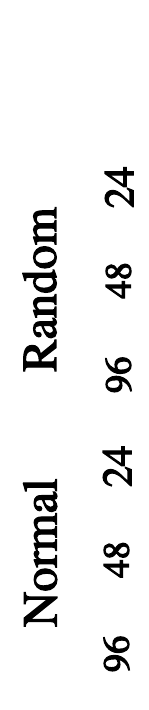}
\end{minipage}
}
\setcounter{subfigure}{0}
\subfloat[MFQ]{
\begin{minipage}{3cm}
\centering
\includegraphics[scale=0.5]{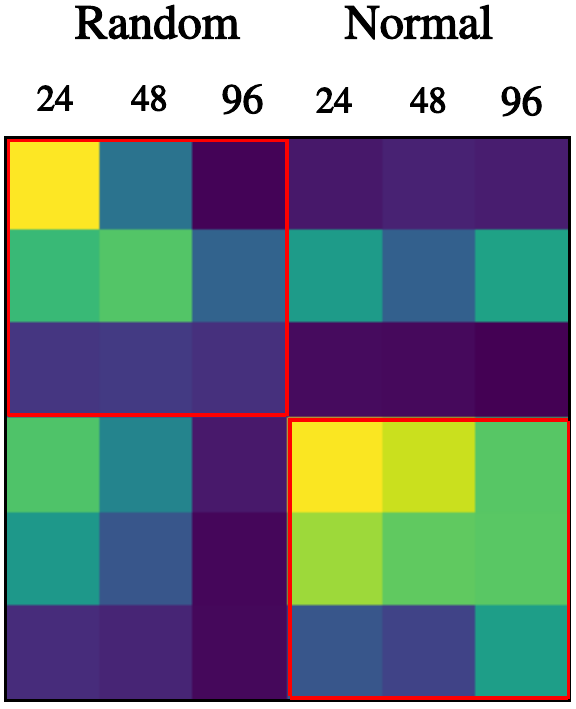}
\end{minipage}
}
\subfloat[NVIF-PPO]{
\begin{minipage}{3cm}
\centering
\includegraphics[scale=0.5]{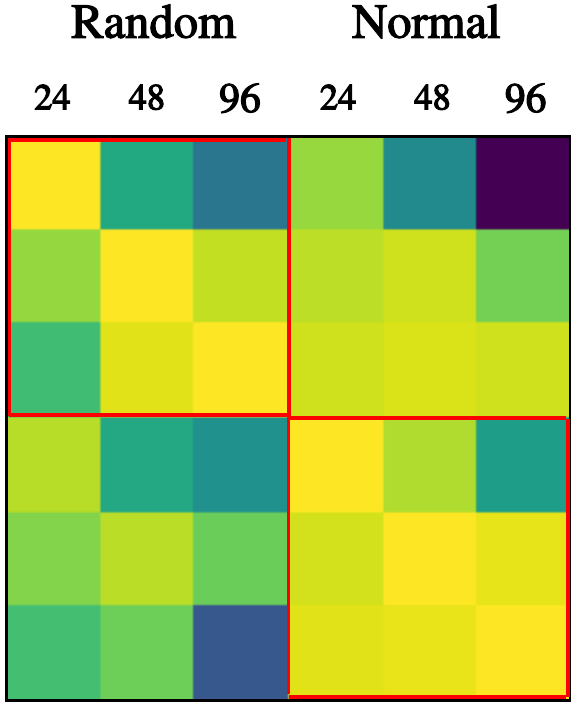}
\end{minipage}
}
\subfloat{
\begin{minipage}{0.5cm}
\centering
\includegraphics[scale=0.45]{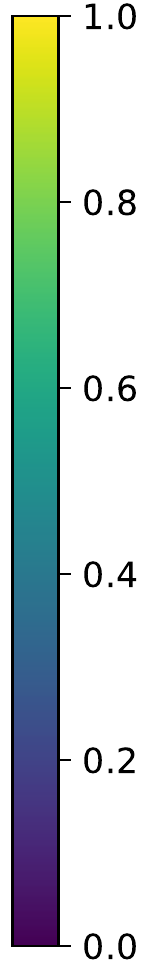}
\end{minipage}
}
\caption{Scalability score matrix of (a) MFQ and (b) NVIF. Scores represent the scalability performance of agents trained in one task when running in another task. }
\label{fig:scale}
\end{figure}

The training of NVIF does not rely on a specific tasks, so that the communication protocol can help the agent policies scale to unseen tasks and achieve good performance. In order to evaluate the scalability of NVIF, we use the policies trained on different maps by NVIF-PPO to conduct 10 episodes on other maps to obtain the average return. The scalability score shown in Fig. \ref{fig:scale} is the average returns normalized by the maximum return in each task. The score at row $i$ and column $j$ indicates that the policy trained in map $i$ is tested in map $j$. Therefore, these scores can represent the scalability of algorithms, with a higher score indicating that the agents can better adapt to other unseen tasks. 

We compare NVIF-PPO with MFQ, which performs better in the experimental algorithms. Since NVIF-PPO can converge to the best performance in all scenarios, the diagonal elements of its score matrix is all 1.0. The closer the color is to yellow, the better the multi-agent system performs in the scalability experiments. As shown in Fig. \ref{fig:scale}, the agents trained by NVIF-PPO have the better scalability than the agents trained by MFQ.

In general, NVIF-PPO can not only achieve good performance in the same task with different map sizes, but also performs well in different task, while MFQ only has scalability in normal tasks. Scores in Fig. \ref{fig:scale} (b) can demonstrate the tendency of agent policies on each maps. In the small-scale maps, the agents tend to complete the tasks without communication and cooperation, so that the policies can hardly scale to larger maps. Besides, since the small maps are more crowded, the policies trained in larger maps are more likely to mistakenly attack other agents. The type of task also affects scalability. Policies trained in normal tasks are hard to be scaled to random tasks because agents do not tend to use communication to promote cooperation. 

\section{Conclusion}
This article proposes NVIF, a novel communication-based MARL method, to improve communication efficiency of large-scale multi-agent system. We adopt neighboring communication to alleviate the problem of information redundancy. The pre-trained NVIF model helps agents to enrich their observations and enhance the stability of the MARL training process. We provide a theoretical analysis of the cooperation in large-scale multi-agent system, which illustrates that the combination of NVIF and PPO can promote cooperation. We also combine NVIF with DQN to test its effectiveness on other RL methods. 

We compare NVIF-PPO and NVIF-DQN with MFQ, DGN, and IPPO in two types of tasks with different map sizes, which are modified from MAgent. Experiments show that NVIF improve the communication efficiency and help agents achieve better performance. We conduct ablation experiments to demonstrate the effect of the neighboring communication and the latent state. We also analyze the learned strategies and conduct supplementary experiments to show the good scalability performance of NVIF.


%

\bibliographystyle{IEEEtran}
\bibliography{IEEEabrv, ref}




\end{document}